\RequirePackage{rotating}
\documentclass[useAMS,usenatbib]{mn2e}
\pdfoutput=1
\usepackage{psfig}
\usepackage{amsmath}
\usepackage{amssymb}
\usepackage{upgreek}
\usepackage{longtable}
\usepackage[mathscr]{eucal}
\usepackage{appendix}
\usepackage{graphicx}
\usepackage{float}
\usepackage{afterpage}
\usepackage{subcaption}
\usepackage[export]{adjustbox}
\usepackage[usenames, dvipsnames]{color}
\usepackage{titlesec}
\usepackage{csvsimple}
\titlespacing*{\section}{0pt}{1.1\baselineskip}{\baselineskip}
\titlespacing*{\subsection}{0pt}{1.1\baselineskip}{\baselineskip}

\title{The Structure and 3D Kinematics of Vela OB2}
\author
[Armstrong et al.]
{Joseph J. Armstrong$^{1,2}$, Nicholas J. Wright$^2$, R.D. Jeffries$^2$, R.J. Jackson$^2$ and \newauthor Tristan Cantat-Gaudin$^3$\\
$^{1}$Department of Space, Earth \& Environment, Chalmers University of Technology, SE-412 96 Gothenburg, Sweden \\
$^{2}$Astrophysics Group, Keele University, Keele, ST5 5BG, UK \\
$^{3}$Max-Planck-Institut für Astronomie, Königstuhl 17, D-69117 Heidelberg, Germany
}
\date{July 2022}

\begin{document}
\maketitle

\begin{abstract}
The kinematics of stars in OB associations can provide insights into their formation, dynamical evolution, and eventual fate. The low-mass stellar content of OB associations are sufficiently numerous as to provide a detailed sampling of their kinematic properties, however spectroscopy is required to confirm the youth of individual stars and to get 3D kinematics. 
In this paper we present and analyse results from a large spectroscopic survey of Vela OB2 conducted using 2dF/HERMES on the AAT. This spectroscopy is used to confirm the youth of candidate young stars and determine radial velocities, which are combined with proper motions and parallaxes from {\it Gaia} to measure 3-dimensional positions and velocities. 
We identify multiple separate kinematic groups in the region, for which we measure velocity dispersions and infer their virial states. We measure expansion rates for all these groups and find strong evidence for anisotropic expansion in the Vela OB2 association of at least 11$\sigma$ significance in all three dimensions, as well as some evidence for expansion in the $\gamma$ Vel and P Puppis clusters. 
We trace back the motions of these groups into the past and find that the open cluster NGC 2547 is an interloper in the Vela OB2 region and actually formed $>$100 pc away from the association. 
We conclude that Vela OB2 must have formed with considerable spatial and kinematic substructure over a timescale of $\sim$10 Myr, with clear temporal substructure within the association, but no clear evidence for an age gradient.
\end{abstract}


\section{Introduction}

OB associations are gravitationally unbound groups of young stars that share an associated motion through space. They are so named after their brightest members, OB stars, the grouping and kinematics of which have long been a subject of interest \citep{ambartsumian47,blaauw64,dezeeuw99}. However, the low-mass members of these associations have only more recently begun to be identified and their complex substructures revealed \citep{preibisch02,briceno07,cantatgaudin19a,zari19}. 

The origins of OB associations are still debated. It has been suggested that they could be the remnants of young clusters which became unbound through the process of residual gas expulsion and began to expand \citep{tutukov78,hills80,kroupa01b}. Once massive O and B type stars have formed in a young cluster their stellar winds and photoionizing radiation expel the molecular gas surrounding them and so the cluster loses the majority of its binding mass and becomes unbound. This hypothesis postulates that OB associations are an intermediate stage between bound clusters and their dispersal into the Galactic field. 

However, more recent investigations have found that groups of young stars form with considerable spatial substructure \citep{sanchez09,tobin2009,andre10,wright14}, and with the availability of high precision astrometry from {\it Gaia} \citep{gaia16,gaiadr2summary}, kinematic substructure has also been identified in OB associations \citep{wright16,wright18,armstrong20}. The spatial and kinematic substructure has since been linked to age substructure as well \citep{cantatgaudin19a,zari19,damiani19}. This evidence better supports a scenario where stars form in subgroups at a range of densities, at different times and with different kinematics \citep{kruijssen12,ward18}. The densest regions of young stars may remain as bound clusters, but the unbound majority form associations which expand and eventually disperse.

Vela OB2 is a nearby (411 $\pm$ 12 pc; \citealt{dezeeuw99}) intermediate age (10 - 20 Myr; \citealt{sahu92,cantatgaudin19a}) association which spans an area of $\sim$100 square degrees on the sky \citep{dezeeuw99,armstrong18,cantatgaudin19a}. The seminal study of OB associations using \textit{Hipparcos} astrometry by \citet{dezeeuw99} identified 93 O-B members of Vela OB2, though until the release of Gaia data \citep{gaia16,gaiadr2summary} the full extent of the low-mass population was unknown.

Vela OB2 is known to contain the young open cluster $\gamma$ Vel whose pre-main sequence (PMS) members have been identified with X-ray \citep{pozzo00,jeffries09} and spectroscopic observations \citep{jeffries14,armstrong20}. Using the spectroscopic sample of $\gamma$ Vel from the Gaia-ESO Survey (GES;\citealt{gilmore12}), \citet{jeffries14} identified two populations of young PMS stars, offset in their radial velocities (RVs) by 2.15 $\pm$ 0.48 km s$^{-1}$. Population A has a narrow RV dispersion $\sigma_A$ = 0.34 $\pm$ 0.16 km s$^{-1}$ and population B has a broader RV dispersion $\sigma_B$ = 1.60 $\pm$ 0.37 km s$^{-1}$, suggesting that population A correlates to the likely bound $\gamma$ Vel cluster and population B to the sparse Vela OB2 association. Indeed, \citet{sacco15} also identified a sparse population towards NGC 2547, $\sim$2 degrees south of $\gamma$ Vel, which exhibits similar RVs to that of \citet{jeffries14}.

The wealth of high-precision astrometry available from {\it Gaia} has facilitated many recent studies investigating the large-scale structure of Vela OB2 and the dynamics of its clusters \citep{damiani17,armstrong18,beccari18,franciosini18,cantatgaudin19a,cantatgaudin19b,armstrong20}. \citet{armstrong18} combined Gaia DR1 and 2MASS \citep{cutri03} photometry to identify the extended PMS population of Vela OB2 and detected considerable substructure outside the $\gamma$ Vel cluster. This structure was also confirmed by \citet{beccari18} and \citet{cantatgaudin19a} who both used {\it Gaia} DR2 and different clustering algorithms to identify multiple groups of young stars that were distinct in position and proper motion. \citet{cantatgaudin19b} extended this work over a wider area around Vela and Puppis to identify seven distinct populations that differ in age and kinematics. The youngest and most densely populated of these corresponds to Vela OB2 and the $\gamma$ Vel cluster, as well as other substructures identified by \citet{armstrong18} and \citet{beccari18}.

In \citet{armstrong20} we presented the results of spectroscopic observations centered on the $\gamma$ Vel cluster. We calculated equivalent-widths of the Li $6707.8 \AA\ $ line (EW(Li)s), a youth indicator for low-mass PMS stars, and RVs for 248 PMS candidates and, combined with the sample of $\gamma$ Vel cluster members from \citet{jeffries14}, separated them into two distinct populations. We searched for expansion trends in three dimensions for both the cluster (population A) and association (population B) components of the sample. We found $>4\sigma$ evidence for expansion in three dimensions for the Vela OB2 association component, though it was significantly anisotropic.

In this study we present results from a larger- and wider-scale spectroscopic survey across the entire Vela OB2 association with the goal of identifying young stars and studying their kinematics. Section \ref{section_data} outlines the data used and the spectroscopic observations performed. Section \ref{section_overview} provides an overview of the region and dissects the young stellar population into different groups, with Section \ref{section_dynamics} going on to analyse the dynamics of these groups. In Section \ref{section_discussion} we discuss our results.

\section{The data}
\label{section_data}

In this section we explain how the sample of young stars studied in this work was compiled, including the spectroscopic data obtained from 2dF/AAT, astrometric data from {\it Gaia}, and photometric data from various sources. We also outline our criteria for cleaning the sample and identifying a reliable list of young stars.

\subsection{Spectroscopic target selection}

\begin{figure}
	\includegraphics[width=\columnwidth]{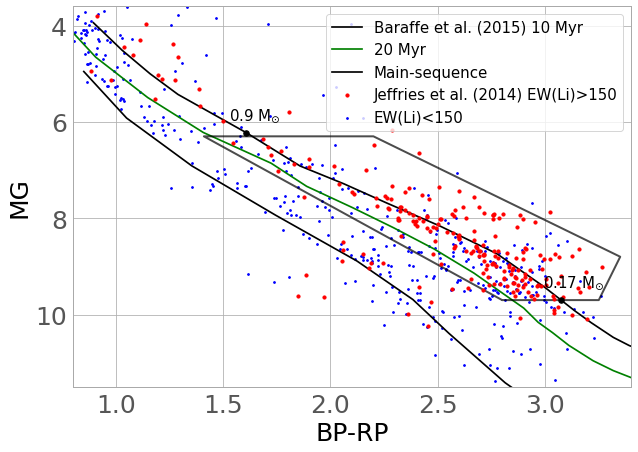}
	\setlength{\belowcaptionskip}{-10pt}
	\setlength{\textfloatsep}{0pt}
	\caption{{\it Gaia} DR2 $G_{BP}-G_{RP}$ vs $M_G$ colour - absolute magnitude diagram showing sources from \protect\citet{jeffries14} with EW(Li)$>150\,$m\AA\ (red) and EW(Li)$<150\,$m\AA\ (blue) using distance estimates from \protect\citet{bailerjones18}. Also shown are 10 and 20 Myr PMS isochrones \protect\citep{baraffe15} after applying a reddening and extinction of $E(V-I)=0.055$ and $A_{V}=0.131$ \protect\citep{jeffries14}. The PMS selection box shown is designed to include the majority of confirmed PMS stars (red) above the 20 Myr \protect\citep{baraffe15} isochrone and within the range $6.5 < M_G < 9.5$. The positions of 0.9 $M_{\odot}$ and 0.17 $M_{\odot}$ PMS stars on the 10 Myr isochrone are shown to indicate the mass range of our targets for spectroscopic observations. }
	\label{DR2cmd}
\end{figure}

We began our selection of likely PMS stars for spectroscopic observation by identifying the regions of Vela OB2 with the highest densities of PMS candidate stars using {\it Gaia} DR2 photometry and parallaxes (Fig. 1). We created a $G_{BP}$ - $G_{RP}$ versus $M_G$ colour-absolute magnitude diagram (CaMD) for {\it Gaia} DR2 sources and designed a selection box to be used as the primary filter, based on the positions of previously identified PMS members of the $\gamma$ Vel cluster and Vela OB2 (\citealt{jeffries14}; Fig. \ref{DR2cmd}). We also employed a parallax cut, removing objects with $\varpi > 4$ or $2 > \varpi$ (sources with distance $>$ 500 pc or $<$ 250 pc respectively), effectively eliminating foreground and background contamination.

\begin{figure*}
	\includegraphics[width=500pt]{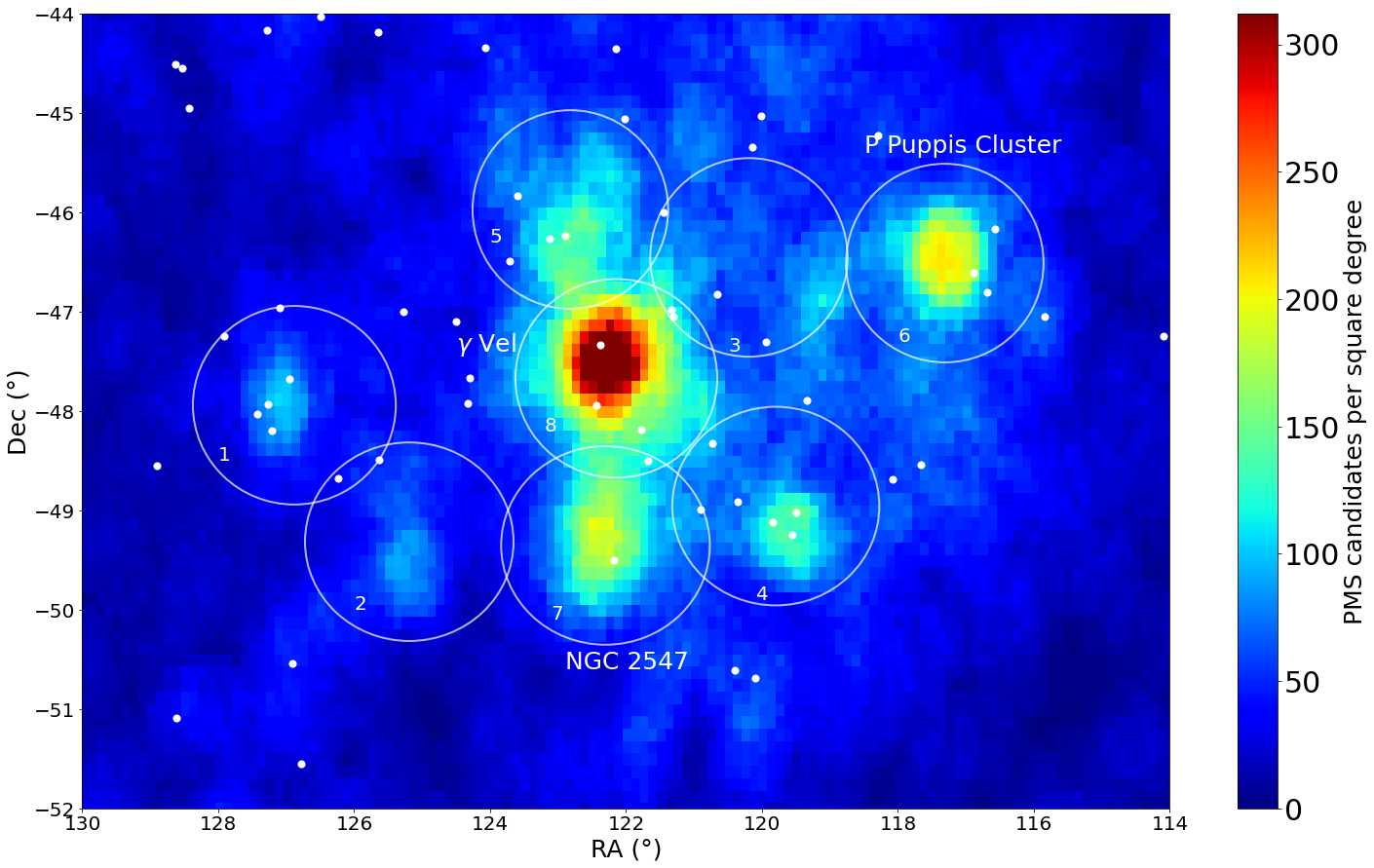}
	\setlength{\belowcaptionskip}{-10pt}
	\setlength{\textfloatsep}{0pt}
	\caption{A density map showing the distribution of likely low-mass PMS stars in the Vela OB2 region, selected using {\it Gaia} DR2 $G_{BP}$, $G_{RP}$ and $G$ band photometry as well as parallaxes (Fig. 2). The known open clusters identified are labeled and massive Vela OB2 members \protect\citep{dezeeuw99} are shown as white dots. Fields chosen for spectroscopic observations are indicated by white circles with field numbers next to them.}
	\label{DR2map}
\end{figure*}

This selection of candidate PMS stars allowed us to produce a map of the distribution of our targets (Fig. \ref{DR2map}). This map clearly shows a wide distribution of candidate PMS stars, as well as the prominent $\gamma$ Vel cluster (RA,Dec = 122.0$^{\circ}$,-47.5$^{\circ}$). Also visible is the open cluster NGC 2547 (RA,Dec = 119.5$^{\circ}$, -49.3$^{\circ}$) and a smaller cluster to the west (RA,Dec = 122.0$^{\circ}$, -46.5$^{\circ}$) that appears to be the P Puppis cluster identified by \citet{caballero08}. Another overdensity is visible at (RA,Dec = 119.5$^{\circ}$, -49.3$^{\circ}$). These clusters correlate well with the groups identified by \citet{beccari18} and are also surrounded by a less dense, widespread population spanning $\sim15\times8$ degrees that broadly follow the distribution of known OB-type members of Vela OB2 \citep{dezeeuw99}.

\subsection{Spectroscopic observations and data reduction}

In order to obtain spectroscopic radial velocities to complement {\it Gaia} 5-parameter astrometry, and spectroscopic youth indicators for low-mass PMS stars (Li and H$\alpha$) we made observations with the 2-degree field (2dF; \citealt{2dF}) fibre positioner and the high-efficiency and resolution multi-element spectrograph (HERMES; \citealt{HERMES}) at the Anglo-Australian Telescope (AAT). 

Using the density map we selected the positions for 8 target fields (white circles in Fig. \ref{DR2map}, listed in Table \ref{field_table}), covering the regions of highest PMS star density and a more diffuse region in field 3. For each field we aimed to select 340 - 350 targets per field to make use of all available fibres on the 2dF/HERMES spectrograph, prioritising targets based on their proximity to 10-20 Myr PMS isochrones (Fig. \ref{DR2cmd}). In total we assigned fibres to 2762 targets in 8 fields, 2635 of which were unique.

\subsubsection{Observations and data reduction}

Observations were made on four nights from 10th - 14th January 2019. Multiple 2400s exposures were taken for each of the 8 selected fields. Coordinates, total exposure time, numbers of targets and numbers of confirmed PMS stars (Section 2.4) for each field are given in Table \ref{field_table}. Calibration frames, including dark frames and multi-fibre flat fields were taken for each field. 25 fibres per field were positioned on regions of empty sky to measure the sky spectrum.

\begin{table*}
\begin{center}
{\renewcommand{\arraystretch}{1.5}
\begin{tabular}{|p{2.5cm}|p{2cm}|p{2cm}|p{2.5cm}|p{2cm}|p{2.5cm}| }
\hline
Field number & RA ($^\circ$) & Dec ($^\circ$) & Exposure time (s) & Targets & Confirmed PMS stars \\
\hline
1 & 126.88001127 & -47.93995199 & 7200 & 344 & 31\\
2 & 125.18927075 & -49.31417082 & 4800 & 349 & 26\\
3 & 120.19231869 & -46.45083941 & 7200 & 342 & 65\\
4 & 119.79694371 & -48.95542601 & 7200 & 348 & 79\\
5 & 122.81934540 & -45.96854395 & 7200 & 343 & 99\\
6 (P Puppis) & 117.31048094 & -46.50807097 & 16800 & 348 & 90\\
7 (NGC 2547) & 122.30222601 & -49.35166480 & 9600 & 340 & 110\\
8 ($\gamma$ Vel) & 122.14277093 & -47.66916442 & 7200 & 348 & 181\\
\hline
Total &  &  &  & 2762 (\textbf{2635}) & 671 (\textbf{653})\\
\hline
\end{tabular}}
\end{center}
\setlength{\belowcaptionskip}{-10pt}
\setlength{\textfloatsep}{0pt}
\caption{Details of the 8 fields targeted and observed, listing the central coordinates, total exposure time (s),  number of science targets and number of spectroscopically confirmed PMS stars per field. As there are some targets included in overlapping fields, the total number of unique targets are given in bold.}
\label{field_table}
\end{table*}

\subsubsection{Data reduction and analysis}

The spectroscopic data were calibrated and reduced using the 2dF Data Reduction (2DFDR) software tool \citep{2dfdr}. Measurement of spectroscopic parameters (equivalent widths of the Li $6707.8 \AA\ $ line and H$\alpha$ 6562.8\AA\ line and H$\alpha$ core and wing indices; EW(Li)s, EW(H$\alpha$)s, $\alpha_{c}$, $\alpha_{w}$) from reduced spectra was done following the procedures of \citet{damiani14}, \citet{jackson18} and \citet{jeffries21}. The measurement of RVs and EW(Li)s required synthetic spectra with matching $T_{\rm eff}$ (derived from G, BP, RP and K band photometry, Section 2.6) to our reduced target spectra. Synthetic spectra were produced using the MOOG spectral synthesis code \citep{Sneden2012a}, with \citet{Kurucz1992a} solar-metallicity model atmospheres, for $\log g =4.5$ and down to $T_{\rm eff} =$ 4000 K in 100k steps. Rotational broadening and instrument resolution are accounted for in the extraction profile.

For sources observed in multiple fields, and where both observations have SNR $<$ 10, we use the mean $\alpha_{w}$ and EW(H$\alpha$) values and weighted mean RVs and EW(Li)s, weighted by the square of the inverse measurement uncertainties.

\subsubsection{Radial velocities}

Reduced target spectra were cross-correlated with their matching synthetic spectra and RVs were determined from the position of the peak in the cross-correlation function (CCF) by fitting a Gaussian function. Sources for which a Gaussian function cannot be satisfactorily fitted to the peak in the CCF (such as spectroscopic binaries) are not given valid RVs and are rejected from our sample.

RV uncertainties were determined empirically from the change in RV between separate exposures of the same target (E$_{RV} = \Delta RV/\sqrt{2}$). The RV uncertainties are normalised per field using a scaling function

\begin{equation}
S_{RV} = FWHM \sqrt{A^2+(B/SNR)^2}
\end{equation}

\noindent where A and B are determined per field as the gradient and intersect of a linear best fit to $1/SNR \ vs \ \Delta RV/\Delta FWHM$ \citep{jeffries21}. Normalised RV uncertainties are then calculated as E$_{RV}$/S$_{RV}$ and these are used in the following analyses.

We obtained RVs and RV uncertainties for $46\%$ of our unique targets.

We also estimate gravitational redshifts for our sample using radii from our SED fits (Section 2.6) and this induced a median offset of $0.52 \pm 0.16$ kms$^{-1}$ in the radial velocities \citep[in good agreement with the findings of][for Pozzo 1, i.e, $\gamma$ Vel]{gutierrez22}.

\subsubsection{Li equivalent-widths} 

The equivalent width of the Li $6707.8 \AA\ $ line (EW(Li)) was measured by subtracting the synthetic spectrum from the target spectrum and then integrating under the relevant profile. EW(Li) uncertainties are taken as the RMS value of the EWs measured using the same procedure with the Gaussian profile of the CCF centred at five wavelengths either side of the Li $6707.8\AA\ $ line \citep{jackson18}. Blended with the Li line is a weak Fe I line at 6707.4 \AA\ which, though the template subtraction should account for this, may mean that EW(Li)s are underestimated by a few m\AA$\,$ if the targets have subsolar metallicities\footnote{\citet{spina17} report median metallicities of $-0.03 \pm 0.02$ dex for $\gamma$ Vel and $-0.006 \pm 0.009$ dex for NGC 2547, consistent with solar metallicity}. 

\begin{figure*}
\includegraphics[width=500pt]{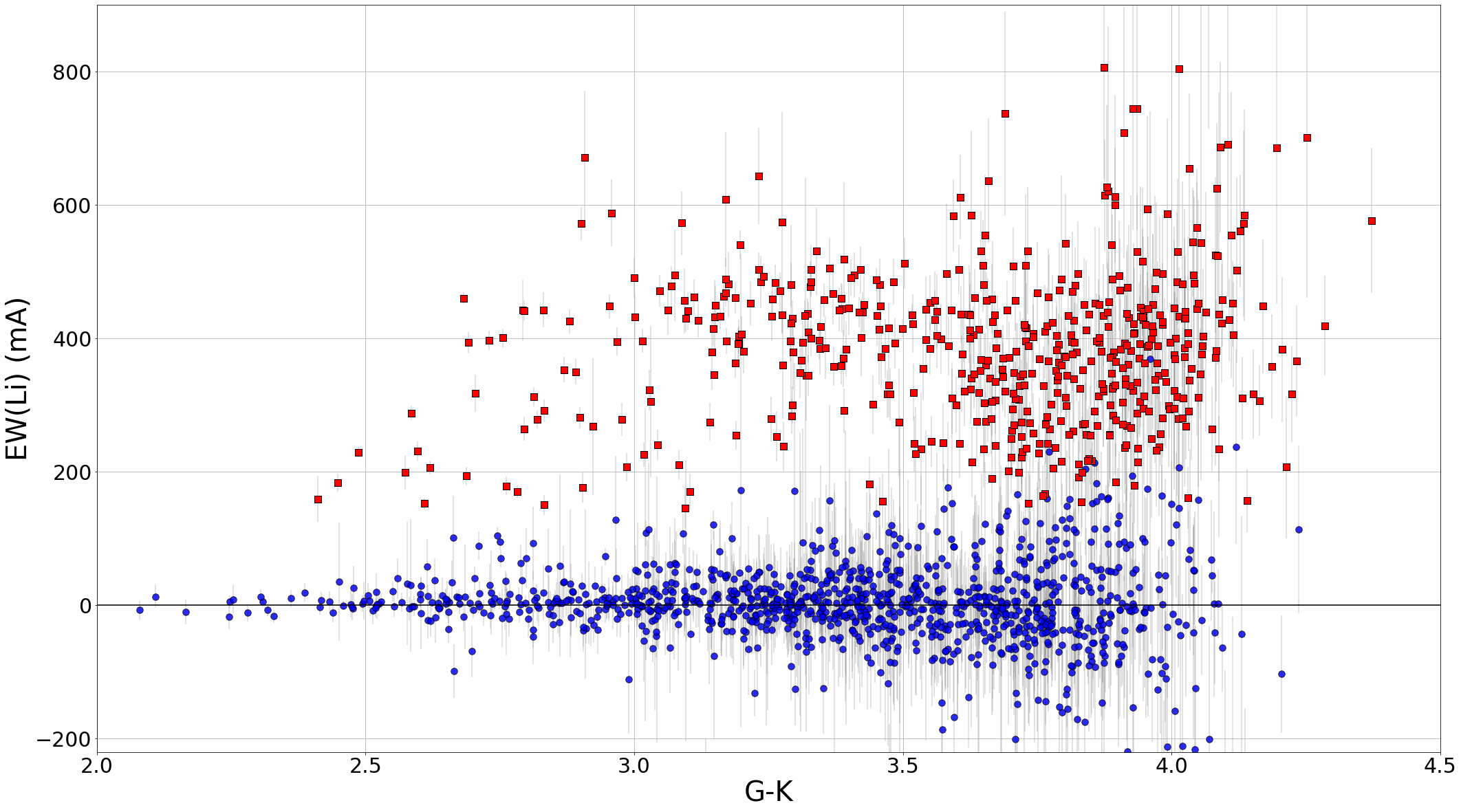}
\setlength{\belowcaptionskip}{-10pt}
\setlength{\textfloatsep}{0pt}
\caption{Equivalent width of the lithium 6708\AA\ line plotted against the $G-K$ colour (combined 2MASS and VHS). Sources that pass our threshold for significant EW(Li) are plotted in red, sources that fail are plotted in blue.}
\label{lithium_selection}
\end{figure*}

As in \citet{armstrong20}, a small zeropoint error in EW(Li) is expected for target stars cooler than 4000 K, the lowest $T_{\rm eff}$ used for the synthetic spectra. However, sources where this is apparent (with negative EW(Li)s) are still consistent with 0 m\AA\ considering their uncertainties ($\sigma_{EW(Li)}$; Fig. \ref{lithium_selection}), and the measured EW(Li)s are sufficiently accurate for the identification of PMS stars.

We obtained EW(Li)s and EW(Li) uncertainties for $74\%$ of our unique targets.

\subsubsection{H$\alpha$}

For stars with high mass accretion rates, excess continuum emission can cause EW(Li)s to be underestimated \citep{palla05}, so other spectroscopic youth indicators are needed to identify such stars. Excess emission of the H$\alpha$ line at 6562.8\AA\ is often used to distinguished between Classical T Tauri stars (CTTs) and Weak-line T Tauri stars (WTTs). \citet{damiani14} define two spectral indices, $\alpha_{c}$ and $\alpha_{w}$, which respectively describe the core and wing components of the H$\alpha$ emission profile. They are defined as

\begin{equation}
\alpha_{c} = \langle f_{core}\rangle/\langle f_{continuum}\rangle
\end{equation}

\begin{equation}
\alpha_{w} = \langle f_{wings}\rangle/\langle f_{continuum}\rangle -0.4\times(1-\langle f_{A2}\rangle/\langle f_{B2}\rangle)
\end{equation}

\noindent where $\langle f_{core}\rangle$ is the mean flux in the wavelength range $6560.8 - 6564.8$\AA\, $\langle f_{wings}\rangle$ is the mean flux in the wavelength ranges $6556.8 - 6560.8$ and $6564.8 - 6568.8$\AA\, $\langle f_{continuum}\rangle$ is the mean flux in the wavelength ranges $6532.8 - 6542.8$ and $6582.8 - 6592.8$\AA\, $\langle f_{A2}\rangle$ is the mean flux in the wavelength range $6530 - 6540$\AA\ and $\langle f_{B2}\rangle$ is the mean flux in the wavelength range $6544 - 6552$\AA\. We note the presence of the [NII] line at 6583\AA\ in many of our spectra and adjust the range of the $f_{continuum}$ components to $6532.8 - 6542.8$ and $6585.8 - 6592.8$\AA\ to exclude it. \citet{damiani14} establish $\alpha_{w}$ $>$ 1.1 as a threshold for candidate CTTs, which we adopt for our selection of PMS stars.

\citet{damiani14} also note that the equivalent width of the H$\alpha$ line can be computed using these indices,

\begin{equation}
EW(H\alpha)(\AA ) = 4\alpha_{c} + 8\alpha_{w} - 12
\end{equation}

\noindent which can also be used to identify candidate CTTs, the commonly adopted threshold for which is EW(H$\alpha$) $>$ 10 m\AA \citep{nikoghosyan19}.

We obtained $\alpha_{c}$, $\alpha_{w}$ and EW(H$\alpha$)s for $99.8\%$, $99.5\%$ and $99.4\%$ of our unique targets.

\subsection{Gaia EDR3 astrometry}

We match our 2635 unique spectroscopic targets to the {\it Gaia} EDR3 catalogue \citep{Gaiaedr3}, improving the precision of 5-parameter astrometry from {\it Gaia} DR2.
In order to ensure the accuracy of our kinematic analysis we require that sources in our sample have a renormalised unit weight error (RUWE) $<$ 1.4 \citep{lindegren18}, large values of which indicate spurious astrometric solutions which may bias any kinematic analysis we perform on them. 1857 of our sources satisfy this requirement, and sources that do not are not included in the subsequent analysis.

Recent analysis of {\it Gaia} EDR3 astrometry have suggested that parallax uncertainties ($\sigma_\varpi$) may be underestimated, especially for bright sources, so we apply a $G$ magnitude dependent correction factor $q$ to $\sigma_\varpi$ as described in equation 16 of \citet{elbadry21}. We also require that sources in our sample satisfy $|\frac{\varpi}{q*\sigma_\varpi}|>2$. All 1857 sources in our sample do satisfy this requirement and their corrected parallax uncertainties are used in the subsequent analysis.

\subsection{Identifying young stars}

For young, low-mass stars Li is a reliable age indicator since it is rapidly depleted once the temperature at the base of the convection zone reaches $3 \times 10^{6}$ K \citep{soderblom10}. The timescale for Li depletion is $\sim$100 Myr for K-type and $\sim$20 Myr for mid M-type stars, making it an effective discriminator between Vela OB2 PMS members and field K- and M- dwarfs (Fig. \ref{lithium_selection}). This is our primary means for identifying PMS stars.

We set the threshold for a significant EW(Li) at EW(Li)$ - \sigma_{EW(Li)} >$ $100\,$m\AA\ as a robust criteria for identifying PMS stars, considering the large $\sigma_{EW(Li)}$ values for redder sources (see Fig. \ref{lithium_selection}). This criteria may exclude a small fraction low-mass members that appear to have already depleted their Li in a narrow temperature range ($3.7 <$ (G-K) $< 4.0$; see also \citealt{jeffries14}), but for our subsequent kinematic analysis excluding these members is preferable to the possible inclusion of older contaminating stars. This may insert a modest bias against older PMS stars, since Li depletion progresses rapidly beyond the age of the $\gamma$ Vel cluster. 624 sources pass this criteria across all 8 fields, 395 of which have valid RVs and RV uncertainties and Gaia EDR3 RUWE$<1.4$. 

We also include in our PMS sample 15 other sources which have valid RVs and RV uncertainties and Gaia EDR3 RUWE$<1.4$ with significant H$\alpha$ excess emission, which satisfy either the $\alpha_{w}$ $>$ 1.1 or the EW(H$\alpha$) $>$ 10 m\AA\ thresholds, but excluding sources with low quality spectra (SNR $<$ 10).

\subsection{Calibration checks}

In our entire sample there are 127 sources with repeat observations in multiple fields, 8 of which meet our spectroscopic youth criteria and have valid RVs and uncertainties from each observation. The standard deviation of the standardised difference in RVs for these sources is 3.226, which we would expect to be $\approx$1 if there was no significant bias between different observations of the same source. This could suggest an additional source of uncertainty exists in our RVs beyond that which we have already quantified, however the number of sources with spectroscopic youth criteria observed multiple times is low and therefore this value is highly uncertain.

To test this we compare the measured RV to the fibre number for all stars in each of our 8 observed fields to search for any dependency that could have introduced a RV bias. We find that the linear best-fit gradients between RV and fibre number are consistent with 0 for 7 fields, while for field 8 ($\gamma$ Vel; Fig. 1) the gradient is only of 1.5$\sigma$ significance.

As a further test we match the sources in our sample to sources observed by \citet{armstrong20}, of which there are 36 young stars with valid RVs and RV uncertainties in both the 2018 and 2019 data. We find that the standard deviation of the standardised difference in RVs for these sources is 0.952, consistent with the expectation value of $\sim$1.

We therefore conclude that there is no significant RV bias or additional RV uncertainty in our measurements and carry out our analysis using the RVs and uncertainties calculated as described above.

\subsection{Estimating stellar parameters with SED fits}

In order to estimate stellar ages and masses for our sources we performed spectral energy distribution (SED) fits to the SEDs for these sources. SEDs were compiled from {\it Gaia}, 2MASS \citep{cutri03} and VISTA Hemisphere Survey (VHS; \citealt{VHS}) photometry\footnote{For sources with $K$-band magnitudes in both catalogues we select the photometry as follows; for $K_{2MASS} < 12$ we use the 2MASS $K$ band, for sources with $K_{VHS} > 13$ we use the VHS $K$ band, otherwise we use the mean of both values. As these sources are nearby and the stellar density of the region is low, we don't expect the difference in resolution of the 2MASS and VHS surveys to introduce any significant bias.}, covering a wavelength range of $\sim 0.4 - 2.2 \mu m$. 1581 sources have at least one infrared band, while 276 sources lack any infrared photometry (though they have {\it Gaia} photometry and so SED fits are still possible).

The SEDs were fit using a forward model and Bayesian inference, with the posterior distribution function sampled using the MCMC sampler {\it emcee} \citep{emcee} for the free parameters of stellar mass and age. The model SEDs were derived using the PARSEC stellar evolution models \citep{marigo17}, which provide effective temperature, luminosity and unreddened photometry. The model SEDs were then reddened by applying a fixed extinction of $A_V = 0.131$ mag ($\gamma$ Vel cluster; \citealt{jeffries14}) and placed at the distance of each source according to their {\it Gaia} EDR3 parallaxes (varying the distance according to the parallax uncertainties). The MCMC sampler was run using 1000 walkers and 500 iterations, with the first half of the posterior distribution discarded as a burn-in and the second half used to derive the best-fit (from the median) and the 1$\sigma$ bounds (from the 16th and 84th percentiles).

The resulting ages and masses may be biased by uncertainties due to unresolved binarity and variability, which are not accounted for in the models. This may introduce a bias in the best-fit ages derived, though the relative ages of the different groups of stars will still be useful. Binarity in particular will mean that observed sources appear more luminous and redder than single stars of the same mass and age, causing their masses to be overestimated and their ages to be underestimated. This will create a tail in the age distribution of sources in our sample extending towards younger ages.

\subsection{Deriving 3D positions and velocities}

We use Bayesian inference to obtain Cartesian positions XYZ and velocities UVW on the Galactic Cartesian system using the coordinate transformation matrices from \citet{johnson87}. To sample the posterior distribution function we use the MCMC sampler \textit{emcee}. For each star we perform 1000 iterations with 100 walkers in an unconstrained parameter space with flat and wide priors (distance priors of 0 - 10 kpc and UVW velocity priors of -200 - 200 kms$^{-1}$). We discard the first half of our iterations as a burn in and from the second half we report the medians of the posterior distribution function as the best fit and use the 16th and 84th percentiles as the 1 $\sigma$ uncertainties (similar to the method used in \citealt{wright18}). Due to the position of Vela OB2 on the sky the Galactic Cartesian Y direction correlates most closely with the line-of-sight and thus has larger uncertainties than X or Z coordinates due to the contribution from parallax uncertainties. Similarly, the uncertainties on Cartesian velocity in the Y direction, V, have a larger contribution from RV uncertainties than velocities U or W.

\subsection{Summary of the data}

After reducing the spectra of our AAT sources we have 1212 unique sources with spectroscopic RV and EW(Li) spread across 8 fields over the Vela OB2 region, with 5-parameter astrometry from Gaia EDR3, with which we calculate Cartesian XYZ positions and UVW velocities. 410 of these sources are identified as PMS stars with significant EW(Li)s, EW(H$\alpha$)s or $\alpha_{w}$s. The median uncertainties on distance, proper motion and RV for the confirmed PMS stars are 6.50 pc, 0.033 mas yr$^{-1}$ (0.06 km s$^{-1}$ at the median distance of 382 pc) and 0.20 km s$^{-1}$ respectively.

\section{Overview of the sample}
\label{section_overview}

\begin{figure*}
	\includegraphics[width=500pt]{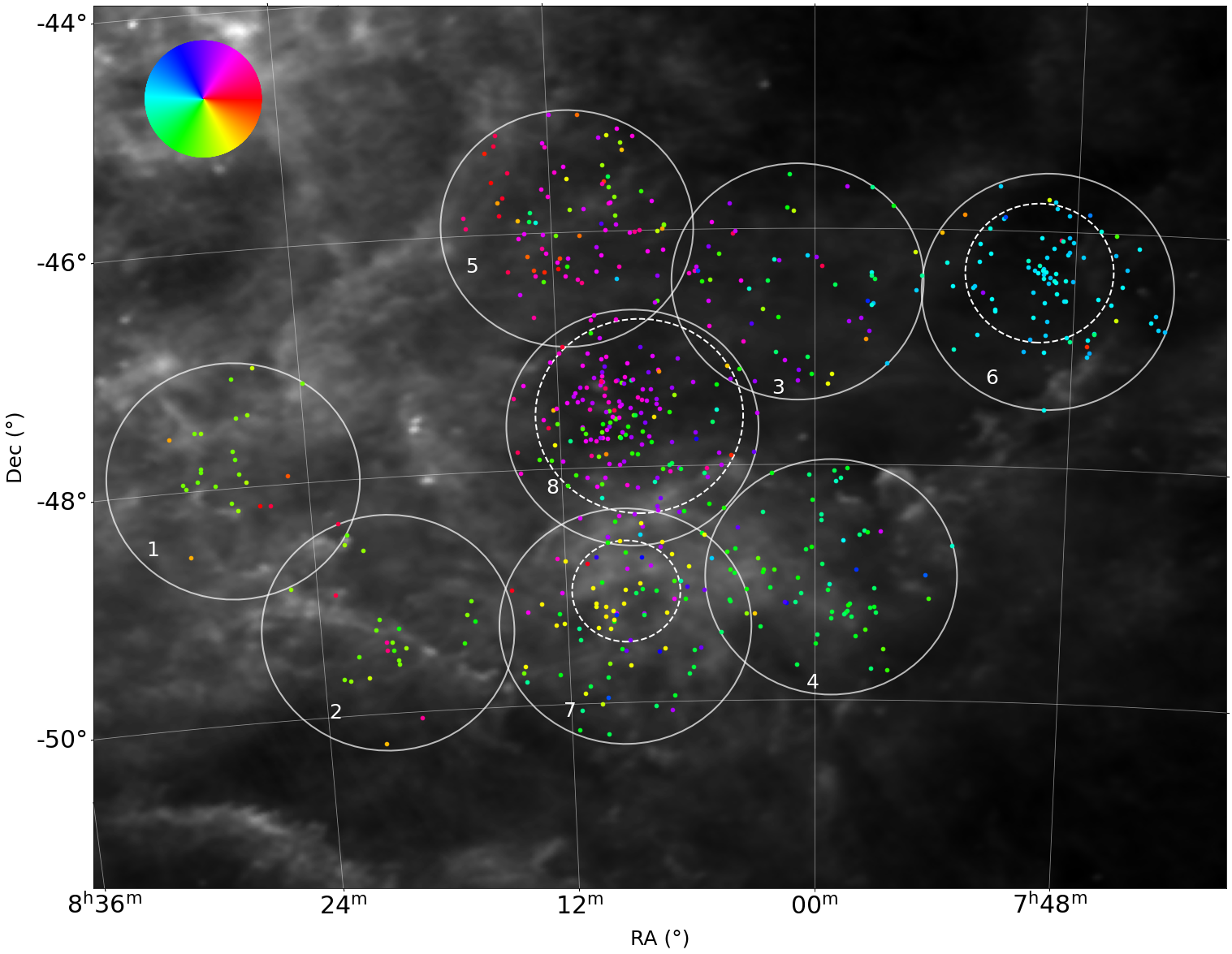}
	\setlength{\belowcaptionskip}{-10pt}
	\setlength{\textfloatsep}{0pt}
	\caption{Spatial distribution of the 410 confirmed PMS stars across the 8 fields observed. Points are colour-coded based on the position angle of the proper motion vector relative to the group median (see the colour wheel in the top left as a key). The background is an IR map of the region from IRIS \citep[Improved Reprocessing of the IRAS Survey,][]{IRIS}. The half-mass radii of the three known clusters ($\gamma$ Vel, NGC 2547 and P Puppis) are indicated by dashed circles.}
	\label{PMS_map}
\end{figure*}

\begin{figure*}
	\includegraphics[width=500pt]{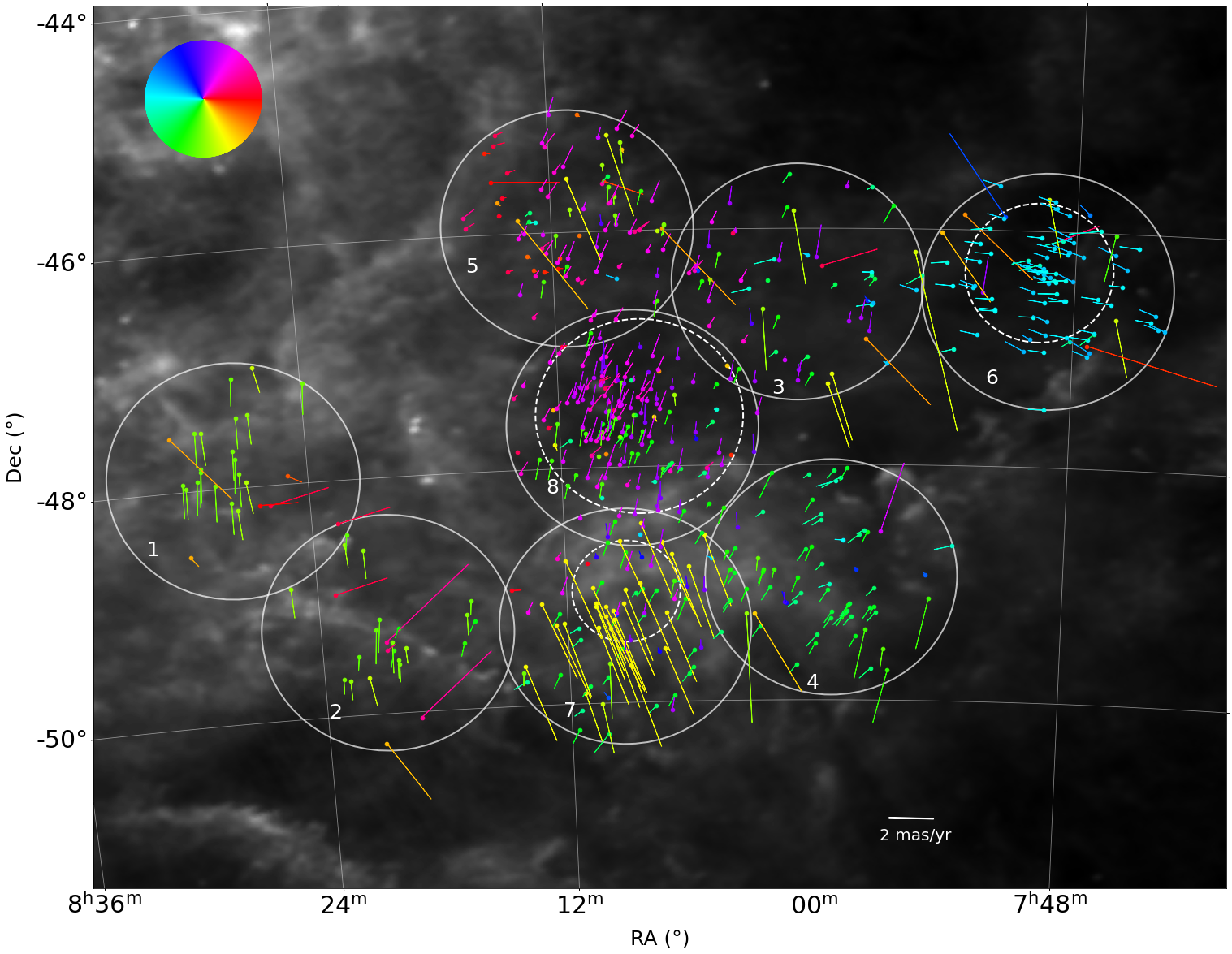}
	\setlength{\belowcaptionskip}{-10pt}
	\setlength{\textfloatsep}{0pt}
	\caption{Spatial distribution of the 410 confirmed PMS stars across the 8 fields observed. Vectors indicate the proper motion of each source relative to the group median, colour-coded based on the position angle of the proper motion (see the colour wheel in the top left as a key). The magnitude scale (mas/yr) of proper motion vectors is indicated by the scale bar in the bottom right. The background is an IR map of the region from IRIS \citep[Improved Reprocessing of the IRAS Survey,][]{IRIS}. The half-mass radii of the three major clusters ($\gamma$ Vel, NGC 2547 and P Puppis) are indicated by dashed circles.}
	\label{PMS_map_pms}
\end{figure*}

Figure \ref{PMS_map} shows the spatial distribution of the 410 confirmed PMS stars with their markers colour-coded according to their relative proper motion (relative to the median of the sample). Figure \ref{PMS_map_pms} shows another version of this figure with the proper motion vectors shown.

Multiple groups of sources are immediately apparent from both the spatial distribution of sources and their proper motions. The dense group of purple-coloured sources at ($8^{h}10^{m}$,$-47.5^{\circ}$) corresponds to the $\gamma$ Vel cluster, the sparser group of yellow-coloured sources at ($8^{h}10^{m}$,$-49^{\circ}$) corresponds to the open cluster NGC 2547 and the group of cyan-coloured sources at ($7^{h}50^{m}$,$-46.5^{\circ}$) is the P Puppis cluster. There are other substructures that are also apparent. While the majority of Vela OB2 sources appear green, with relative motion towards the south-east, a group can be seen in the two eastern-most fields moving southwards and another group in field 5 in red is moving north-west. There are also a number of sources across all fields with proper motions different to the nearest significant groups, some of which might possibly be runaway stars ejected from the clusters in this region.

\begin{figure}
\includegraphics[width=\columnwidth]{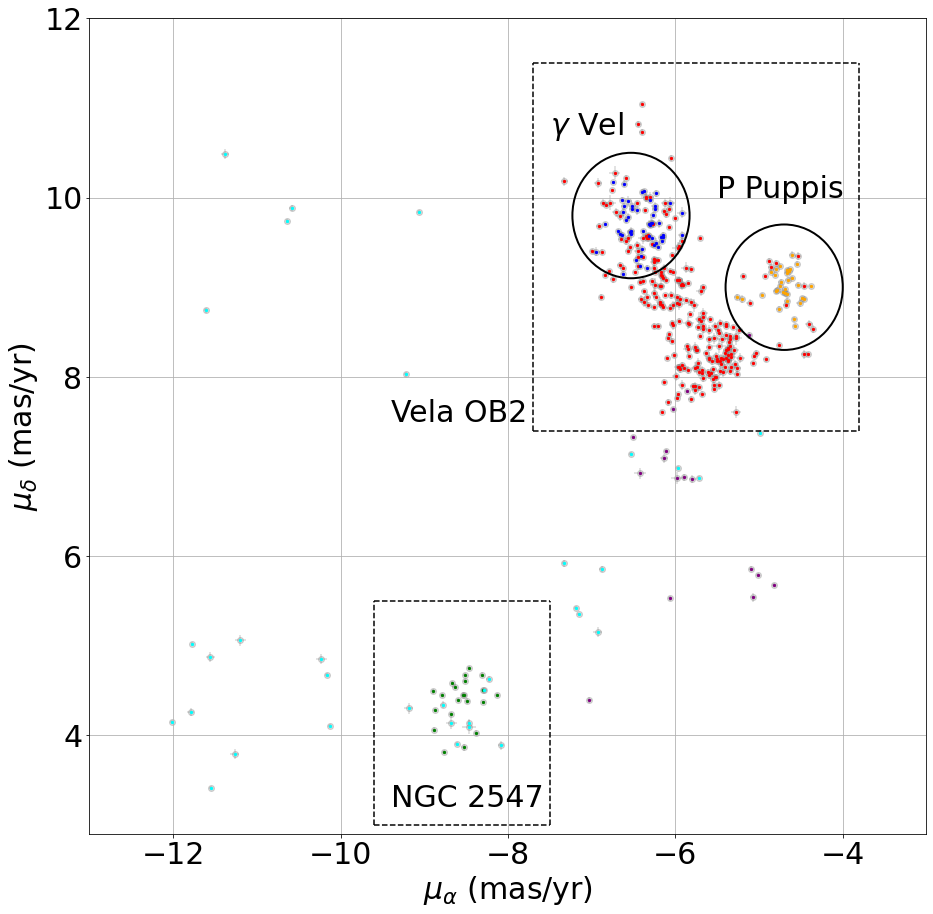}
\setlength{\belowcaptionskip}{-10pt}
\setlength{\textfloatsep}{0pt}
\caption{Proper motions of confirmed PMS stars colour-coded according to their grouping (see Section \ref{s:subgroups}): $\gamma$ Vel cluster (blue), Vela OB2 (red), P Puppis cluster (yellow), NGC 2547 cluster (green), background sources (purple, $d > 440$ pc), with all other sources shown in cyan. The selection areas for the $\gamma$ Vel, P Puppis and NGC 2547 clusters, as well as Vela OB2 are also shown.}
\label{pm_map}
\end{figure}

Figure \ref{pm_map} shows the distribution of these sources in proper motion space. Again, our sample can be divided into distinct groups. We find that the densest region of sources at  ($\mu_{\alpha}, \mu_{\delta}$) $\sim$ (-6.5, 9.5) mas yr$^{-1}$ corresponds to the $\gamma$ Vel cluster, the smaller dense group at ($\mu_{\alpha}, \mu_{\delta}$) $\sim$ (-4.5, 9) mas yr$^{-1}$ is the P Puppis cluster, and the swathe of sources around these two clusters corresponds to the wider Vela OB2 association. It is clear from this that both of these clusters are spatially and kinematically related to Vela OB2. The sources belonging to NGC 2547 are located at ($\mu_{\alpha}, \mu_{\delta}$) $\sim$ (-8.5, 4.5) mas yr$^{-1}$ and appear kinematically distinct from Vela OB2 however.

\begin{figure}
\includegraphics[width=\columnwidth]{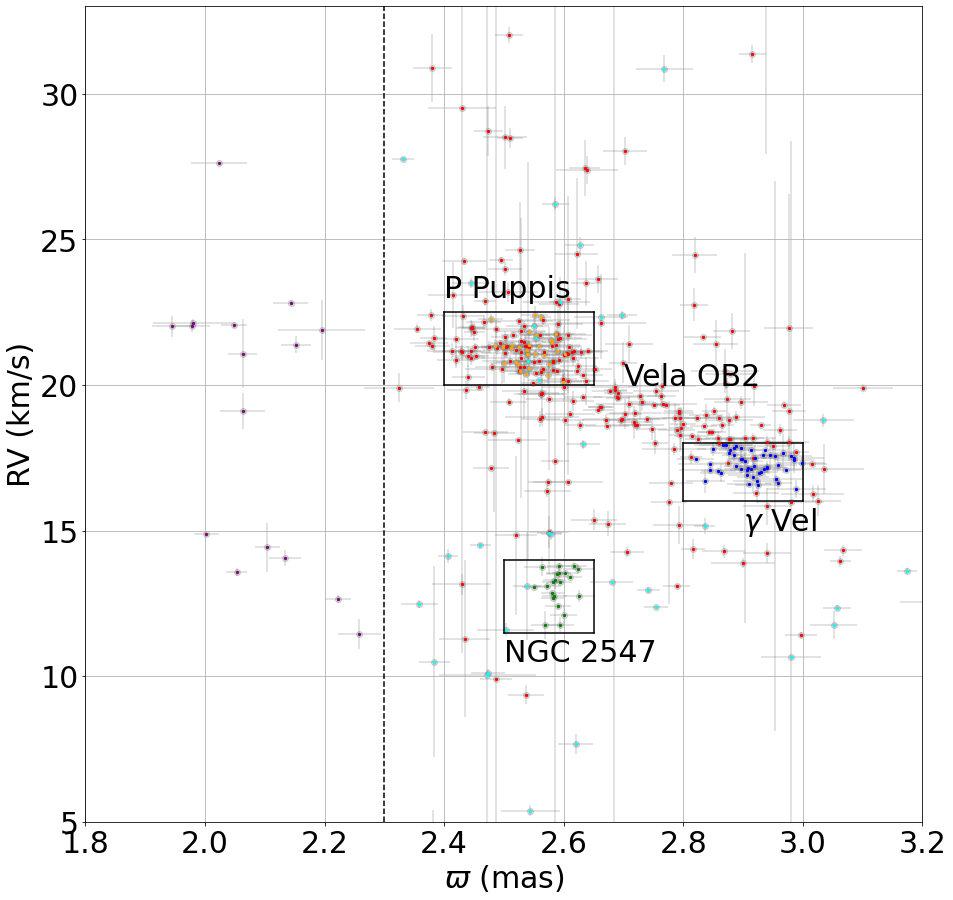}
\setlength{\belowcaptionskip}{-10pt}
\setlength{\textfloatsep}{0pt}
\caption{RV versus parallax for the confirmed PMS stars, with the colour-coding as per Figure \ref{pm_map}. Selection areas for the $\gamma$ Vel, NGC 2547 and P Puppis clusters are shown. The anti-correlation between RV and parallax for Vela OB2 sources is evidence of expansion in the association. The dashed line at $\varpi = 2.3$ mas indicates the boundary between Vela OB2 and the background population.}
\label{rv_parallax}
\end{figure}

In Fig. \ref{rv_parallax} we plot RV versus parallax for our confirmed PMS stars. The $\gamma$ Vel and P Puppis clusters are not as distinct from the Vela OB2 association in this plot, due mostly to the greater uncertainties on both of these quantities, but are visible at around ($\varpi$ = 2.85 mas, RV = 17 kms$^{-1}$) and ($\varpi$ = 2.5 mas, RV = 21 kms$^{-1}$) respectively. NGC 2547 is a distinct group at ($\varpi$ = 2.6 mas, RV = 14 kms$^{-1}$). In addition to the widespread population apparent in this figure we note the presence of a number of stars whose parallaxes are too small ($\varpi < 2.3$ mas) to be a part of the Vela OB2 association, and may represent another group of young stars in the background.

\subsection{Identifying subgroups in the sample}
\label{s:subgroups}

In order to study the properties and kinematics of the different groups of stars in our sample we need to separate them, ideally in a way that does not introduce any kinematic biases. Firstly, we identify the regions of proper motion space occupied by the main clusters (Fig. \ref{pm_map}). We define a selection area for the $\gamma$ Vel cluster of radius 0.7 mas yr$^{-1}$ centred on ($\mu_{\alpha}, \mu_{\delta}$) = (-6.53, 9.8) mas yr$^{-1}$, an area for the P Puppis cluster of radius 0.7 mas yr$^{-1}$ centred at ($\mu_{\alpha}, \mu_{\delta}$) = (-4.7,9) mas yr$^{-1}$ and an area for NGC 2547 within $-9.6 < \mu_{\alpha}$ / mas yr$^{-1} < -7.5$ and $3 < \mu_{\delta}$ / mas yr$^{-1} < 5.5$. We also define boundaries for these clusters in RV and parallax (Fig. \ref{rv_parallax}). For $\gamma$ Vel we require sources to have 2.8 $< \varpi / mas <$ 3.0 and 16 $<$ RV / kms$^{-1}$ $<$ 18, for P Puppis we require sources to have 2.4 $< \varpi / mas <$ 2.65 and 20 $<$ RV / kms$^{-1}$ $<$ 22.5 and for NGC 2547 we require sources to have 2.5 $< \varpi / mas <$ 2.65 and 12 $<$ RV / kms$^{-1}$ $<$ 14.

These boundaries for membership of the clusters are designed to be strict, minimising contamination of the samples at the risk of a reduced completeness. As a consequence it is likely that the Vela OB2 sample will contain sources that are really members of the clusters. However, as it is by far the most populous group, a small amount of contamination will affect our results much less than contamination of the cluster groups would. In figures \ref{PMS_map} and \ref{PMS_map_pms} we show with dashed lines the areas enclosed within each clusters' half-mass radius. These are determined as the radius that contains half of all young stars within these cluster samples.

There is a distinct group of more distant sources at $\varpi <$ 2.25 mas ($d \gtrsim 450$ pc) which we plot in purple in Figures \ref{pm_map} and \ref{rv_parallax}. Sources which are not allocated to the $\gamma$ Vel, P Puppis or NGC 2547 clusters or the distant group but are within the large overdensity in proper motion space ($-7.7 < \mu_{\alpha}$ / mas yr$^{-1} < -3.8$ and $7.4 < \mu_{\delta}$ / mas yr$^{-1} < 11.5$) are allocated to the Vela OB2 association group and are plotted in red in Figures \ref{pm_map} and \ref{rv_parallax}. This is the largest group in our sample. All remaining PMS stars not in these five groups are plotted in cyan in Figures \ref{pm_map} and \ref{rv_parallax}. These sources have kinematics very different from the main groups and may be unrelated to these regions or runaway stars that have been ejected from one of these clusters.

In total we allocate 50 sources to the $\gamma$ Vel cluster, 32 to the P Puppis cluster, 21 to NGC 2547, 246 to Vela OB2, identify 11 sources as distant objects and 26 sources as kinematic outliers. The low number of sources identified as part of NGC 2547 (compared to the number of low-mass members identified by \citealt{jackson22}, for example) is due to the age of this cluster ($\sim$35 Myr, \citealt{jeffries05}) being greater than either the range of ages our CMD selection box was designed for (Fig. \ref{DR2cmd}) and the ages at which many low-mass PMS stars deplete their Li (see Section 2.4).

\subsection{Comparing the sample to other works}

Of the 410 PMS stars we have confirmed, 386 were included by \citet{cantatgaudin19b} as candidate young stars, which we can now confirm. The populations of \citet{cantatgaudin19b} represent groups of young stars in the extended Vela-Puppis region which are distinguished by their different ages and kinematics, but together suggest a prolonged period of connected star formation events from a turbulent molecular cloud. We also note that 418 of our sources with measured EW(Li)s which fail our spectroscopic PMS criteria match to candidate young stars of \citet{cantatgaudin19b}.

Overall, the membership of sources in the Vela OB2 association and the three clusters studied here agrees very well with the populations of \citet{cantatgaudin19b}. From the matches between our sample and the populations of \citet{cantatgaudin19b} all the sources in the $\gamma$ Vel and P Puppis clusters and all but one source in the Vela OB2 association match to population 7 of \citet{cantatgaudin19b}, their youngest population (10-15 Myr). All our sources in NGC 2547 and the majority of our kinematic outliers match to their population 4, an older population (35-40 Myr), suggesting that most of our kinematic outliers may not be related to the young stars of Vela OB2 and its associated clusters. Sources in our 'distant' group are shared between several populations of \citet{cantatgaudin19b}. Most belong to their population 6, a young population associated with the cluster BH 23 (located outside our area of observation), but some also belong to populations 5 and 7 of \citet{cantatgaudin19b}.

\citet{beccari18} applied the DBSCAN clustering algorithm \citep{ester96} to Gaia DR2 sources in this region and identified 6 clusters, of which their clusters 4,3 and 1 correspond to $\gamma$ Vel, NGC 2547 and P Puppis respectively. They estimated isochronal ages of 10 Myr for $\gamma$ Vel and P Puppis and 30 Myr for NGC 2547 using OmegaCAM photometry. 

\citet{pang21} applied the StarGo algorithm \citep{yuan18} to Gaia EDR3 sources in the region and identified 5 kinematic groups, of which their groups Huluwa 1 and 3 correspond to $\gamma$ Vel and P Puppis respectively. They estimated isochronal ages of 12.1 - 22.4 Myr for $\gamma$ Vel and 10.6 - 19.6 Myr for P Puppis using Gaia EDR3 photometry.

\subsection{Ages of the subgroups}
\label{section_ages}

\begin{figure}
	\includegraphics[width=\columnwidth]{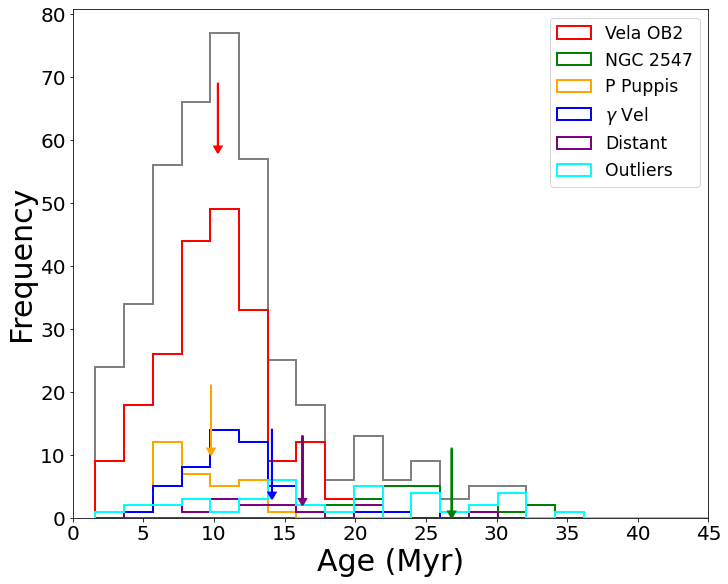} 
	\setlength{\belowcaptionskip}{-10pt}
	\setlength{\textfloatsep}{0pt}
	\caption{Distribution of ages of confirmed PMS stars in each group. Histogram colours indicate the populations of sources using the same colour-coding as Fig \ref{pm_map}, with the grey histogram indicating the total per bin. Weighted mean ages for each group are indicated by coloured arrows. The weighted mean ages are 10.3 Myr for the Vela OB2 group (red), 26.8 Myr for NGC 2547 (green), 9.8 Myr for P Puppis (yellow), 14.1 Myr for $\gamma$ Vel (blue) and 16.3 Myr for the distant population (purple).}
	\label{ages}
\end{figure}

In figure \ref{ages} we show the distribution of ages derived for our confirmed PMS stars from their SED fits. The weighted mean values are 10.3 Myr for the Vela OB2 group, 26.8 Myr for NGC 2547, 9.8 Myr for P Puppis, 14.1 Myr for $\gamma$ Vel and 16.3 Myr for the distant population. We do not report uncertainties on these SED ages since there are many factors which could contribute bias which are not modelled. These ages are in reasonable agreement with, but systematically lower than, the literature ages for these groups \citep[e.g.,][]{sahu92,jeffries05,jeffries14,jeffries17,cantatgaudin19b}. This under-estimation is probably due to a combination of factors, that our photometric target selection is biased against stars older than previously identified members of $\gamma$ Vel (Section 2.1), the fact that our SED fits do not account for binarity (which will make stars appear more luminous and therefore younger) and the inaccuracy of some evolutionary models that do not account for radius inflation in young, low-mass stars \citep{jeffries17}, though the PARSEC stellar evolution models used to derive model SEDs are calibrated to match isochrones of young low-mass stars \citep{chen14} and so produce older ages than other conventional isochrones. 

Scaling up our mean ages by 25-30\% brings them into good agreement with literature ages and allows us to estimate ages for the P Puppis cluster ($\sim$12.5 Myr) and the distant population ($\sim$20 Myr) that are on the same scale as the literature ages for these groups. Regardless of the scaling it is notable that while $\gamma$ Vel and Vela OB2 have similar ages, the P Puppis cluster is distinctly younger, indicating a timescale of star formation in this region spanning up to $\sim$10 Myr assuming the age scale used here.

The kinematic outliers have similar ages to the stars of NGC 2547, as well as having similar proper motions (Fig. \ref{pm_map}) and belonging to population 4 of \citet{cantatgaudin19b}. However, we do not include them in NGC 2547 due to their different parallaxes and RVs (Fig. \ref{rv_parallax}).

\section{Dynamics of the subgroups}
\label{section_dynamics}

Now that we have identified a population of young stars across Vela and separated these stars into multiple groups and clusters we can use these samples to study the dynamics of these groups.

\subsection{Velocity dispersions}

\begin{figure} 
    \subfloat{{\includegraphics[width=240pt]{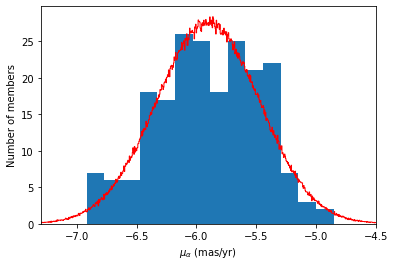} }}%
    \qquad
    \subfloat{{\includegraphics[width=240pt]{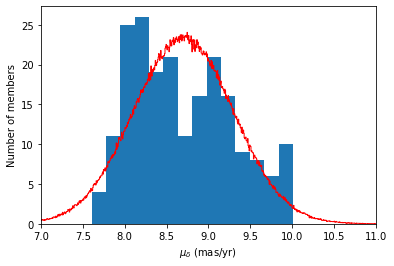} }}%
    \qquad
    \subfloat{{\includegraphics[width=240pt]{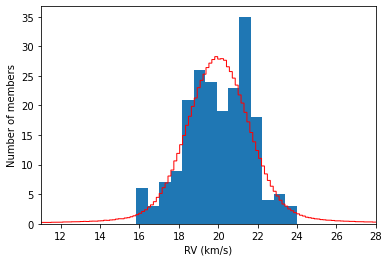} }}%
    \setlength{\belowcaptionskip}{-10pt}
    \setlength{\textfloatsep}{0pt}
    \caption{Histograms of observed velocities for sources belonging to the Vela OB2 association group (blue). Model distributions (red) are produced by sampling our best fitting models 1,000,000 times with additional uncertainties randomly sampled from the observed velocity uncertainties.}
    \label{Contamination}%
\end{figure}

Velocity dispersions can be a useful indication of the dynamical state of a group of stars and its gravitational boundedness. We estimate the velocity dispersions for each group and cluster in our sample using Bayesian inference, sampling the posterior distribution with a Markov Chain Monte Carlo (MCMC) sampler and comparing the observations to the model using a maximum likelihood \citep[see e.g.,][]{wright19}.

The model velocity distributions are 3-dimensional Gaussians with a total of 6 free parameters (the central velocity and velocity dispersion in each dimension). For each star modelled we then add an uncertainty sampled from the observed uncertainty distribution in each dimension.

To model the effects of unresolved binarity on our RV distribution we add instantaneous velocity offsets to the modelled RVs for a fraction of the modelled stars equal to the binary fraction, which we take to be 46\% \citep{raghavan10}. The velocity offsets are taken from a population of modelled binaries with primary star masses between $0.1 - 0.65$ M$_{\odot}$ (to match the range of our observed sample) using a \citet{maschberger13} IMF. Secondary star masses are sampled between 0.1--1.0 of their primary star mass with a uniform probability. The distribution of orbital periods is log-normal with mean period $log_{10}(5.03)$ and dispersion $log_{10}(2.28)$ days \citep{raghavan10}. The distribution of eccentricities is flat between $e=0$ to a maximum that scales with the orbital period \citep{parker09}. We randomise the inclination of binaries in 3D relative to the line of sight and choose random times in the binary orbits to calculate velocities along the line of sight of each star. We then apply a luminosity weighting between the velocities of each star in the binary. We don't consider triple systems, as the contribution of the third star to the observed RV is likely to be small, and the properties of these systems are not well constrained.

We sample the posterior distribution function using the MCMC ensemble sampler \textit{emcee} and use an unbinned maximum likelihood test to compare the model and observations. We use wide and uniform priors for each free parameter, of $-100$ to $+100$ kms$^{-1}$ for central velocities and 0 to $100$ kms$^{-1}$ for velocity dispersions. We use 1000 walkers and perform 2000 iterations, the first half of which is discarded as burn-in. We take the median value of the posterior distribution as the best fit and the 16th and 84th percentiles as 1$\sigma$ uncertainties.

Table~\ref{kinematic_table} lists the best fit central velocities and velocity dispersions for all the groups and clusters studied. Figure~\ref{Contamination} shows the 3D velocity distributions for stars in Vela OB2 as an example, with the best-fitting velocity dispersion models overplotted. The best fitting velocity dispersions for Vela OB2 are significantly anisotropic, with $\sigma_{\mu_{\delta}}$ (km/s) = $0.786^{+0.047}_{-0.039}$ and $\sigma_{RV}$ (km/s) = $1.424^{+0.153}_{-0.133}$ implying anisotropy with a confidence of $4.5\sigma$. As OB associations are sparse and unbound they are believed to be dynamically un-evolved \citep{wright16} and so retain their initial substructure, which is indicated by velocity anisotropy.

The best fitting velocity disperions for the P Puppis cluster also show evidence of anisotropy ($\sigma_{\mu_{\alpha}}$ (km/s) = $0.340^{+0.068}_{-0.051}$ and $\sigma_{\mu_{\delta}}$ (km/s) = $0.200^{+0.043}_{-0.032}$), albeit of only $2\sigma$ significance. This could suggest that the P Puppis cluster has not yet undergone sufficient dynamical mixing to develop isotropy, though more precise measurements are needed to confirm this.

Both the $\gamma$ Vel and NGC 2547 clusters have velocity dispersions consistent with being isotropic at the 1.5 and $<$1 sigma levels. This suggests that both of these clusters are sufficiently mixed to have erased any primordial anisotropy they may have possessed. This is not particularly surprising for NGC~2547 given its age as it would be expected to be reasonably well mixed. The velocity dispersions in radial velocity for $\gamma$ Vel ($\sigma_{RV}$ (km/s) = $0.310^{+0.085}_{-0.074}$) and for Vela OB2 ($\sigma_{RV}$ (km/s) = $1.424^{+0.153}_{-0.133}$) are in good agreement with the velocity dispersions calculated by \citet{jeffries14} for their concentrated ($\sigma_{RV}$ (km/s) = $0.34 \pm 0.16$) and dispersed populations ($\sigma_{RV}$ (km/s) = $1.6 \pm 0.37$).

We calculate 3D velocity dispersions, which are also listed in Table~\ref{kinematic_table}. We calculate virial masses for each group and cluster according to
\begin{equation}
    M_{vir}=\eta \frac{\sigma_{3D}^2 r_{eff}}{3G}
\end{equation}
\noindent where $\eta = 10$, and $r_{eff}$ is the radius within which half of a group or cluster's members are located. These are given in Table \ref{kinematic_table}. 

The virial mass estimated for Vela OB2 ($7073^{+1212}_{-1048}$ M$_{\odot}$) is significantly larger than the total stellar mass of 1285$\pm$110 M$_{\odot}$ estimated by \citet{armstrong18}, as expected for an unbound OB association. The virial masses for the $\gamma$ Vel, NGC 2547 and P Puppis clusters are likely to be more consistent with their stellar masses as the long-lived nature of these clusters, coupled with their broadly isotropic velocity dispersions, suggests that they are gravitationally bound. These are also broadly consistent with literature estimates; $370 M\odot$ \citep{littlefair03} and $450 \pm 100 M\odot$ \citep{jeffries04} for NGC 2547, $152 M\odot$ \citep{jeffries14} for $\gamma$ Vel.

\begin{table*}
\begin{center}
{\renewcommand{\arraystretch}{1.5}
\begin{tabular}{|p{3.5cm}|p{2.5cm}|p{2.5cm}|p{2.5cm}|p{2.5cm}| }
\hline
 & Vela OB2 & $\gamma$ Vel & NGC 2547 & P Puppis \\
\hline
Median $\mu_{\alpha}$ (mas/yr) & -5.92 $\pm$ 0.43 & -6.39 $\pm$ 0.18 & -8.52 $\pm$ 0.22 & -4.67 $\pm$ 0.10 \\
Median $\mu_{\delta}$ (mas/yr) & 8.60 $\pm$ 0.59 & 9.71 $\pm$ 0.23 & 4.47 $\pm$ 0.17 & 9.00 $\pm$ 0.17 \\
Median RV (km/s) & 20.03 $\pm$ 1.65 & 17.28 $\pm$ 0.40 & 13.23 $\pm$ 0.40 & 21.29 $\pm$ 0.33 \\
Median Distance (pc) & $382.8^{+0.8}_{-0.7}$ & $345.6^{+1.1}_{-1.1}$ & $386.2^{+1.3}_{-1.2}$ & $391.8^{+1.8}_{-1.8}$ \\
Central Position X (pc) & $-45.8^{+0.1}_{-0.1}$ & $-41.8^{+0.2}_{-0.2}$ & $-38.4^{+0.2}_{-0.2}$ & $-66.0^{+0.4}_{-0.4}$ \\
Central Position Y (pc) & $-376.0^{+0.8}_{-0.8}$ & $-339.5^{+1.2}_{-1.1}$ & $-379.8^{+1.2}_{-1.2}$ & $-380.0^{+2.0}_{-1.8}$ \\
Central Position Z (pc) & $-53.4^{+0.2}_{-0.2}$ & $-48.1^{+0.3}_{-0.3}$ & $-57.7^{+0.2}_{-0.2}$ & $-68.6^{+0.4}_{-0.4}$ \\
Central Velocity U (km/s) & $-21.21^{+0.04}_{-0.03}$ & $-21.03^{+0.08}_{-0.07}$ & $-16.34^{+0.06}_{-0.06}$ & $-22.22^{+0.11}_{-0.12}$ \\
Central Velocity V (km/s) & $-17.32^{+0.10}_{-0.10}$ & $-14.53^{+0.05}_{-0.06}$ & $-10.04^{+0.06}_{-0.05}$ & $-17.45^{+0.05}_{-0.05}$ \\
Central Velocity W (km/s) & $-3.61^{+0.02}_{-0.02}$ & $-2.88^{+0.03}_{-0.02}$ & $-10.85^{+0.05}_{-0.04}$ & $-3.13^{+0.02}_{-0.02}$ \\
$\sigma_{\mu_{\alpha}}$ (mas/yr) & $0.595^{+0.037}_{-0.031}$ & $0.240^{+0.031}_{-0.027}$ & $0.191^{+0.050}_{-0.036}$ & $0.183^{+0.037}_{-0.027}$ \\
$\sigma_{\mu_{\alpha}}$ (km/s) & $1.080^{+0.067}_{-0.057}$ & $0.393^{+0.051}_{-0.044}$ & $0.350^{+0.091}_{-0.066}$ & $0.340^{+0.068}_{-0.051}$ \\
$\sigma_{\mu_{\delta}}$ (mas/yr) & $0.433^{+0.026}_{-0.022}$ & $0.185^{+0.027}_{-0.021}$ & $0.249^{+0.065}_{-0.046}$ & $0.108^{+0.023}_{-0.017}$ \\
$\sigma_{\mu_{\delta}}$ (km/s) & $0.786^{+0.047}_{-0.039}$ & $0.303^{+0.044}_{-0.034}$ & $0.456^{+0.119}_{-0.084}$ & $0.200^{+0.043}_{-0.032}$ \\
$\sigma_{RV}$ (km/s) & $1.424^{+0.153}_{-0.133}$ & $0.310^{+0.085}_{-0.074}$ & $0.426^{+0.199}_{-0.130}$ & $0.255^{+0.121}_{-0.112}$ \\
$\sigma_{3D}$ (km/s) & $1.952^{+0.167}_{-0.145}$ & $0.586^{+0.102}_{-0.086}$ & $0.716^{+0.239}_{-0.163}$ & $0.469^{+0.133}_{-0.112}$ \\
Virial mass (M$_{\odot}$) & $7073^{+1212}_{-1048}$ 
& $272^{+95}_{-80}$ & $285^{+190}_{-130}$ & $133^{+75}_{-63}$ \\
\hline
\end{tabular}}
\end{center}
\setlength{\belowcaptionskip}{-10pt}
\setlength{\textfloatsep}{0pt}
\caption{Kinematic properties for Vela OB2, NGC 2547, P Puppis and the $\gamma$ Vel cluster. See the text for a discussion of how these quantities were derived.}
\label{kinematic_table}
\end{table*}

\subsection{Expansion}
\label{section_expansion}

To determine whether any of the groups or clusters are expanding we search for correlations between velocity and position in each dimension. Positive or negative correlations between positions and velocity in the same dimension indicate either expansion or contraction respectively. We do this in the Galactic Cartesian coordinate system $XYZ$ and fit a linear relationship between position and velocity.  To minimise the effect of outliers we remove $>3\sigma$ outliers in position and velocity before conducting the fits. We determine the best fitting parameters of this relationship using Bayesian inference and explore the posterior distribution using MCMC.

We model the gradient and intersect of the linear fit and the fractional amount by which the uncertainties are underestimated ($m$, $b$, $f$). We assume that errors follow a Gaussian distribution and are independent, and use linear least squares for maximum likelihood estimation. The likelihood function is given as

\begin{equation}
ln p(y|x,\sigma,m,b,f) = -\frac{1}{2}\sum_{n}[\frac{(y_{n}-mx_{n}-b)^{2}}{s_{n}^{2}}+ln(2\pi s_{n}^{2})]
\end{equation}

\noindent where

\begin{equation}
s_{n}^{2}=\sigma_{n}^{2}+f^{2}(mx_{n}+b)^{2}
\end{equation}

\noindent and where $\sigma_n$ are velocity uncertainties for the $n$ data points and $f$ quantifies under-estimated measurement or model uncertainties \citep{hogg10}. Uncertainties in position are accounted for by varying the measured position according to it's uncertainties during the MCMC simulation. This is repeated for 2000 iterations with 200 walkers, half of which are discarded as burn in, the second half from which medians and 16th and 84th percentiles are reported from the posterior distribution function as the linear best fit gradient and uncertainties. 

\begin{table}
\begin{center}
{\renewcommand{\arraystretch}{1.5}
\begin{tabular}{|p{1.3cm}|p{1cm}|p{1cm}|p{2.0cm}|p{0.8cm}| }
\hline
Group & Velocity & Position & Gradient & Signif. \\
 & & & (km s$^{-1}$ pc$^{-1}$) \\
\hline
Vela OB2 & $U$ & $X$ & $ 0.074_{-0.005}^{+0.005}$ & 14$\sigma$\\
         & $V$ & $Y$ & $ 0.046_{-0.004}^{+0.004}$ & 11$\sigma$ \\
         & $W$ & $Z$ & $ 0.050_{-0.004}^{+0.004}$ & 12$\sigma$ \\
\hline
P Puppis & $U$ & $X$ & $ 0.094_{-0.026}^{+0.025}$ & 3$\sigma$ \\
         & $V$ & $Y$ & $ 0.003_{-0.025}^{+0.029}$ & - \\
         & $W$ & $Z$ & $ 0.030_{-0.015}^{+0.014}$ & 2$\sigma$ \\
\hline
$\gamma$ Vel & $U$ & $X$ & $ 0.062_{-0.013}^{+0.012}$ & 4$\sigma$ \\
         & $V$ & $Y$ & $ 0.007_{-0.012}^{+0.012}$ & - \\
         & $W$ & $Z$ & $ 0.034_{-0.012}^{+0.011}$ & 2$\sigma$ \\
\hline
\end{tabular}}
\end{center}
\setlength{\belowcaptionskip}{-10pt}
\setlength{\textfloatsep}{0pt}
\caption{Expansion gradients fitted for each group and in each dimension. Significance values listed are calculated from the ratio of the gradient to the uncertainty on the gradient, rounded down to the nearest integer.}
\label{expansion_table}
\end{table}

\begin{figure*}
\begin{subfigure}{0.33\textwidth}
    \includegraphics[width=165pt]{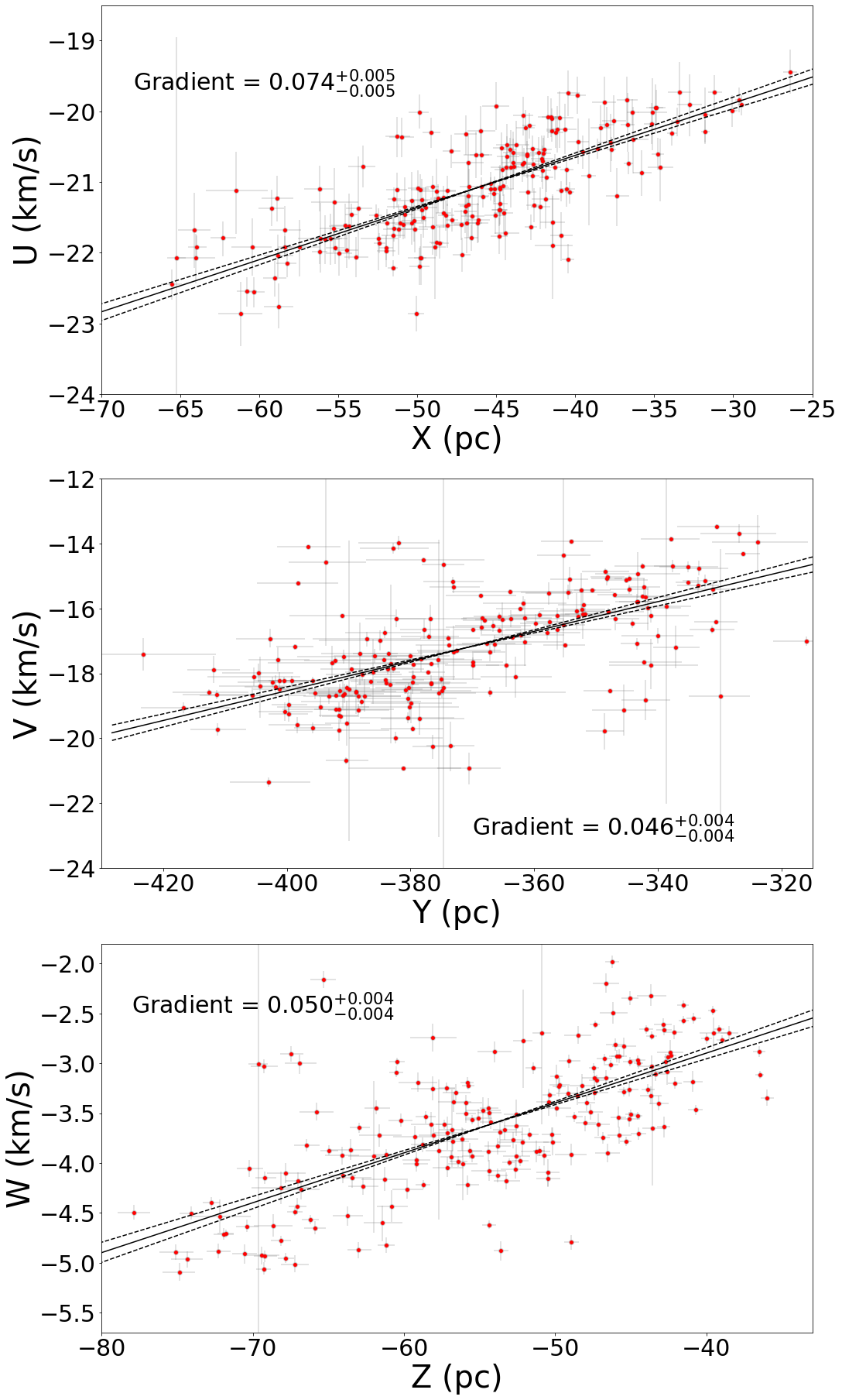}
\end{subfigure}
\begin{subfigure}{0.33\textwidth}
    \includegraphics[width=165pt]{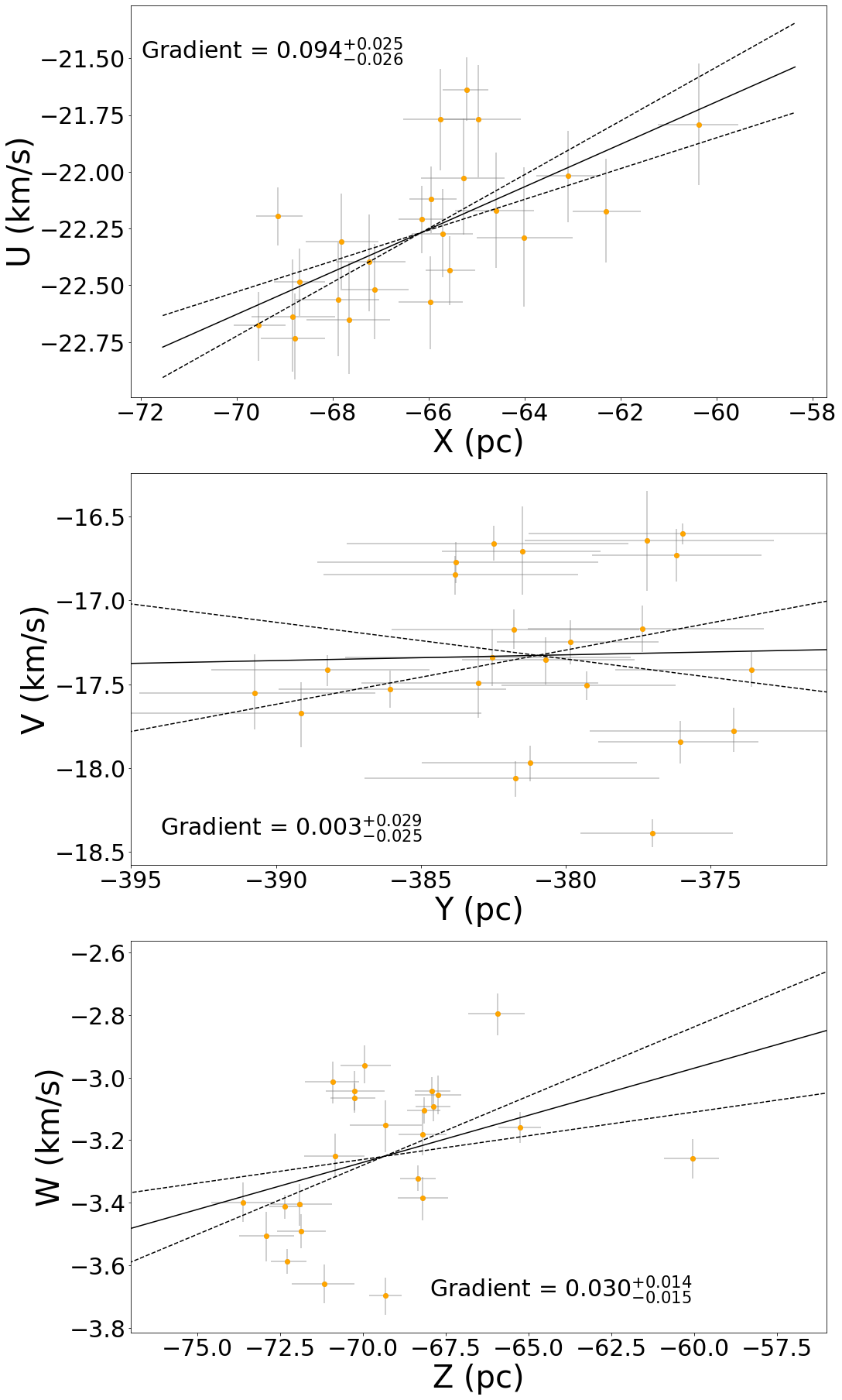}
\end{subfigure}
\begin{subfigure}{0.33\textwidth}
    \includegraphics[width=165pt]{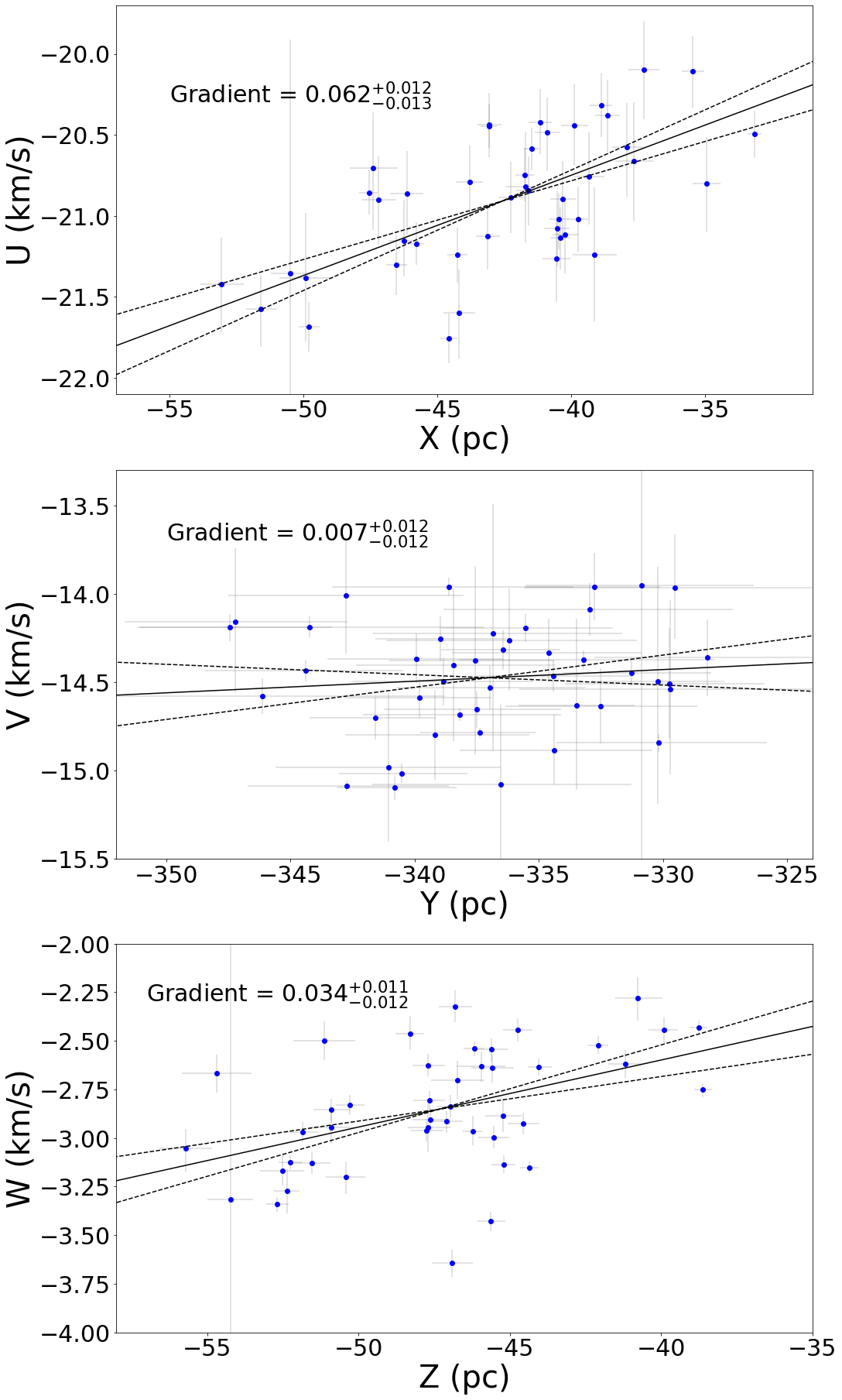}
\end{subfigure}
\caption{
Cartesian velocity versus position in each of the three dimensions $XYZ$ for stars in the Vela OB2 association (red, left), the P Puppis cluster (yellow, middle) and the $\gamma$ Vel cluster population (blue, right) with uncertainties shown. The best-fitting gradients and the 16th and 84th percentiles values of the fit are shown as solid and dashed lines respectively in each panel.}
\label{figure_expansion}
\end{figure*}

The best fitting gradients are listed in Table~\ref{expansion_table} and are shown in Figure~\ref{figure_expansion} for position vs expansion component of velocity. We find very strong evidence for expansion in Vela OB2, with a strong correlation between position and velocity in each dimension that we have measured with significances of 11--14 $\sigma$. 
The expansion is strongest in $X$ vs $U$, which is similar to the findings of \citet{armstrong20} who also found a strong level of anisotropic expansion in Vela OB2, though the signatures of expansion they found were less significant than ours. 

There is some evidence for expansion in the P Puppis and $\gamma$ Vel clusters, both of which show significant expansion in the $X$ direction (at 3 and 4 $\sigma$, respectively), some evidence of expansion in the $Z$ direction (2$\sigma$), but no significant expansion in the $Y$ direction. There is no significant evidence for expansion in NGC~2547, due mostly to the small number of sources observed in this cluster.

Given that both the P Puppis and $\gamma$ Vel clusters show their strongest expansion in $X$ vs $U$, the same direction in which Vela OB2 exhibits expansion, there is a risk that contamination from the latter to the former has contributed to the measured expansion in the clusters. Our membership criteria for the compact clusters was intentionally conservative, to minimise such effects, but it may still be an issue. Comparison of the distribution of Vela OB2 sources in Figure~\ref{figure_expansion} with the distribution of $\gamma$ Vel sources shows they overlap heavily, particularly in the $U$ vs $X$ plot (see Figure~\ref{Contamination}), and therefore the strong gradient measured in this dimension may be due to contamination from Vela OB2. Figure~\ref{Contamination} also compares the distribution of P Puppis sources in $X$ vs $U$ compared to the distribution of Vela OB2 sources. The two groups are clearly separated, which indicates that the P Puppis group is unlikely to be highly contaminated by Vela OB2 sources, and that the expansion gradients for this group do represent its true physical state.

\begin{figure} 
    \subfloat{{\includegraphics[width=225pt]{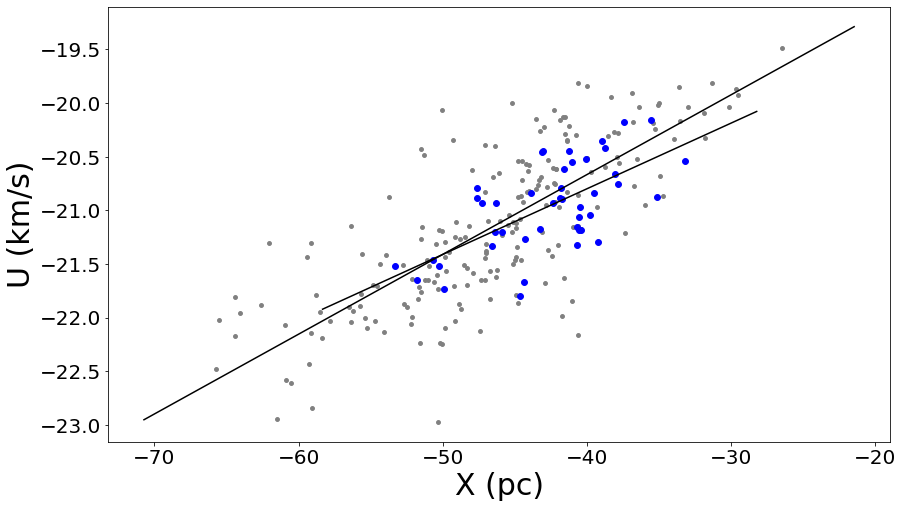} }}%
    \qquad
    \subfloat{{\includegraphics[width=225pt]{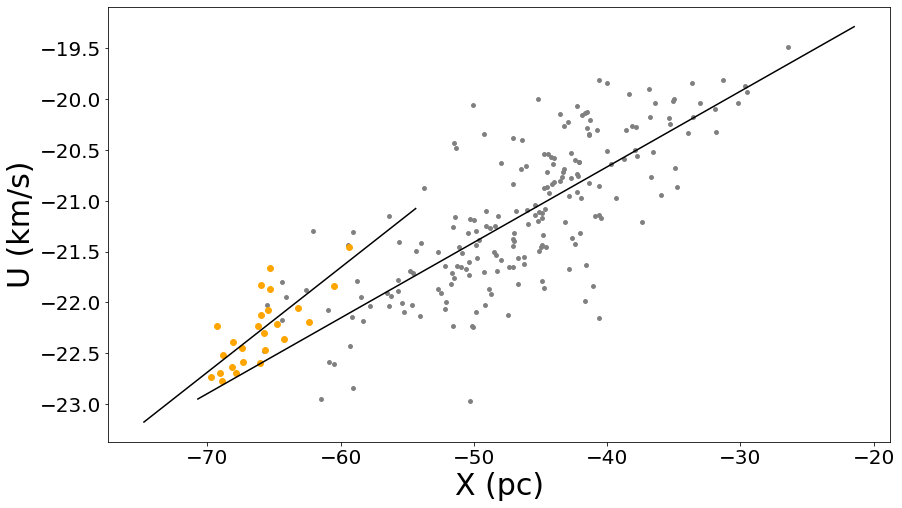} }}%
    \setlength{\belowcaptionskip}{-10pt}
    \setlength{\textfloatsep}{0pt}
    \caption{$Top:$ $U$ vs $X$ for sources in $\gamma$ Vel (blue) and Vela OB2 (grey) with MCMC linear best fits shown as solid lines. $Bottom:$ $U$ vs $X$ for sources in the P Puppis cluster (yellow) and in Vela OB2 (grey) with MCMC linear best fits shown as solid lines.}
    \label{Contamination}
\end{figure}

\subsection{Rotation}
\label{rotation}

Correlations between position and velocity in different dimensions can provide an indication of rotation in a group of stars. We repeat the same gradient fits performed in Section \ref{section_expansion} between position and velocity in different dimensions to search for evidence of rotation in our groups. 

\begin{table}
\begin{center}
{\renewcommand{\arraystretch}{1.5}
\begin{tabular}{|p{1.3cm}|p{1cm}|p{1cm}|p{1.7cm}|p{1.3cm}| }
\hline
Group & Velocity & Position & Gradient (kms$^{-1}$/pc) & Signif. \\
\hline
Vela OB2 & $V$ & $X$ & $ -0.004_{-0.015}^{+0.016}$ & - \\
 & $W$ & $X$ & $ 0.003_{-0.006}^{+0.006}$ & - \\
 & $U$ & $Y$ & $ 0.001_{-0.002}^{+0.002}$ & - \\
 & $W$ & $Y$ & $ 0.019_{-0.002}^{+0.001}$ & 9$\sigma$ \\
 & $U$ & $Z$ & $ -0.015_{-0.005}^{+0.005}$ & 2$\sigma$ \\
 & $V$ & $Z$ & $ 0.076_{-0.011}^{+0.011}$ & 6$\sigma$ \\
\hline
P Puppis & $V$ & $X$ & $ 0.052_{-0.054}^{+0.050}$ & - \\
 & $W$ & $X$ & $ -0.033_{-0.023}^{+0.028}$ & 1$\sigma$ \\
 & $U$ & $Y$ & $ 0.023_{-0.015}^{+0.012}$ & 1$\sigma$ \\
 & $W$ & $Y$ & $ -0.011_{-0.012}^{+0.012}$ & 1$\sigma$ \\
 & $U$ & $Z$ & $ 0.002_{-0.027}^{+0.029}$ & - \\
 & $V$ & $Z$ & $ 0.021_{-0.041}^{+0.042}$ & - \\
\hline
$\gamma$ Vel & $V$ & $X$ & $ -0.023_{-0.011}^{+0.012}$ & 1$\sigma$ \\
 & $W$ & $X$ & $ -0.014_{-0.011}^{+0.012}$ & 1$\sigma$ \\
 & $U$ & $Y$ & $ 0.002_{-0.015}^{+0.015}$ & - \\
 & $W$ & $Y$ & $ 0.018_{-0.009}^{+0.009}$ & 1$\sigma$ \\
 & $U$ & $Z$ & $ -0.035_{-0.017}^{+0.018}$ & 1$\sigma$ \\
 & $V$ & $Z$ & $ 0.017_{-0.014}^{+0.013}$ & 1$\sigma$ \\
\hline
\end{tabular}}
\end{center}
\setlength{\belowcaptionskip}{-10pt}
\setlength{\textfloatsep}{0pt}
\caption{Rotation gradients for Vela OB2, the P Puppis and $\gamma$ Vel clusters derived from fitting linear gradients between position and velocity in different dimension.}
\label{table_rotation}
\end{table}

Table~\ref{table_rotation} lists the best fit rotation gradients for Vela OB2 and the P Puppis and $\gamma$ Vel clusters (NGC 2547 is too sparsely sampled to provide reliable rotation fits). In the two clusters we find no significant evidence for expansion in any dimension. However, in Vela OB2 there is strong evidence for expansion between $W$ and $Y$ and between $V$ and $Z$, of 9 and 6 $\sigma$, respectively. 

\begin{figure}
	\includegraphics[width=\columnwidth]{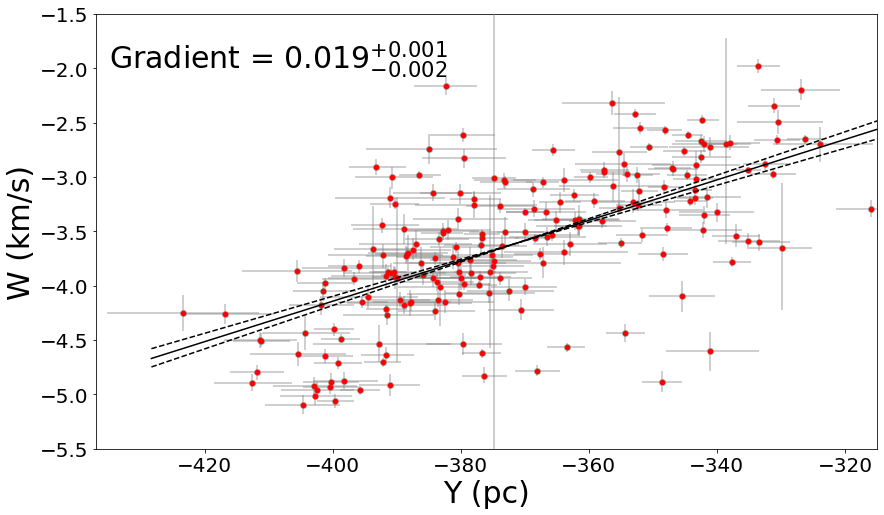} 
	\setlength{\belowcaptionskip}{-10pt}
	\setlength{\textfloatsep}{0pt}
	\caption{$Y$ vs $W$ rotation of Vela OB2 sources (red) with MCMC linear best fit and 16th and 84th percentiles shown as solid and dashed lines respectively. }
	\label{figure_rotation}
\end{figure}

The former is shown in Figure~\ref{figure_rotation} illustrating the strong correlation between velocity, $W$, and position, $Y$. Given that the association is gravitationally unbound, these trends more likely represent residual angular momentum in the dynamics of this system, rather than rotation.

\subsection{Estimating kinematic ages}

If a group of stars is expanding, and one assumes that the stars were originally in a very compact configuration (such as a star cluster) then the expansion gradient can be used to estimate the timescale for the expansion. This is known as the kinematic age and is derived from the reciprocal of the expansion gradient in each dimension. We derived kinematic ages in each dimension for Vela OB2, the only one of our groups to be conclusively shown to be expanding. The ages are $13.8_{-0.9}^{+1.0}$ Myr in $X$, $22.2_{-1.8}^{+2.1}$ Myr in $Y$ and $20.5_{-1.5}^{+1.8}$ Myr in $Z$. The anisotropic expansion observed clearly gives unequal kinematic ages, implying that the Vela OB2 association did not expand from a completely compact (cluster-like) configuration. The kinematic ages of 13-24 Myr are broadly consistent with the literature age of 16-20 Myr \citep{sahu92}, suggesting that the expansion of Vela OB2 may have begun at, or close to, the time of its formation, and that it has been expanding ever since. However, it is older than the age suggested by the fit of low-mass members to PMS isochrones \citep{jeffries17,cantatgaudin19b} our SED age estimates, and the age of $\gamma^2$ Vel \citep[3.5 Myr;][]{north07}, again hinting at the inaccuracy of some evolutionary models (Section 3.3).

\subsection{Past and future structure of the Vela region}

The three clusters $\gamma$ Vel, P Puppis and NGC 2547 have very different kinematics not just from each other, but also from the wider Vela OB2 association. To analyse the past and future structure of the Vela region we can use the 3D positions and velocities to trace back the motions of stars and clusters to study their distribution in the past (including at their birth) and in the future. 

We calculate 3D positions as a function of time using the epicycle approximation and the orbital equations from \citet{fuchs06}. We use the Oort A and B constants from \citet{feast97}, the local disc density from \citet{holmberg04}, the local standard of rest velocity from \citet{schonrich10} and a solar $Z$ distance above the Galactic plane of 17 pc \citep{karim17}. We perform this traceback on the individual stars in Vela OB2, the distant group and the outlier group. For the clusters $\gamma$ Vel, P Puppis and NGC 2547 we perform the traceback on the entire clusters, on the assumption that they are (or are close to) gravitationally bound. Note that our traceback does not take into account dynamical interactions between stars and investigations using N-body simulations would be needed to probe the past dynamics in more detail.

\begin{figure*}
	\includegraphics[width=490pt]{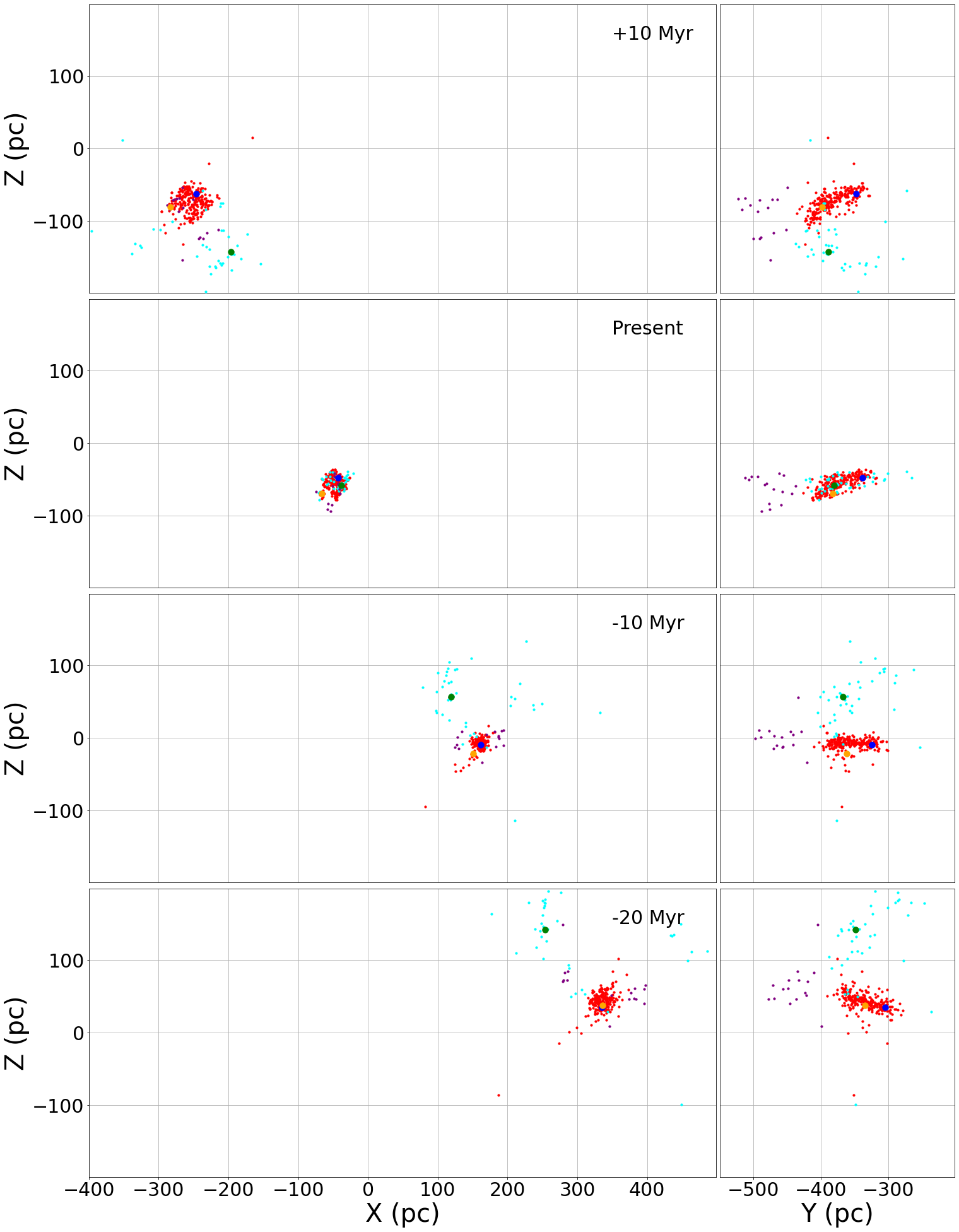} 
	\setlength{\belowcaptionskip}{-10pt}
	\setlength{\textfloatsep}{0pt}
	\caption{Spatial distribution of young stars in the Vela region at 10 Myr in the future, at the current time and at 10 and 20 Myr in the past. The $\gamma$ Vel (blue), P Puppis (yellow),  and NGC 2547 clusters (green) are shown as single points, on the assumption that the clusters are gravitationally bound. The positions of sources in Vela OB2 (red), the distant group (purple) and the outlier group (cyan) are shown individually.}
	\label{traceback_wide}
\end{figure*}

Figure~\ref{traceback_wide} shows the spatial distribution of sources and clusters in the Vela over the last 20~Myr. It is clear that NGC 2547 is an interloper to the Vela OB2 region and is just projected against the association at the present time. The position of NGC 2547 is outside the volume of Vela OB2 5 Myrs ago and is $>100$pc away at 20 Myrs in the past. Similar motion is seen for stars in the `outlier group', a group that we noted had stars of a similar age to NGC 2547 (Section \ref{section_ages}). This implies that the formation of these stars may be closely linked to the formation of NGC 2547 itself.

The $\gamma$ Vel and P Puppis clusters remain within the volume of the Vela OB2 association up to 20 Myrs backwards in time, which includes their estimated ages, therefore implying that they formed within the Vela OB2 association. Their future motion suggests the two clusters will continue to move apart, following the overall expansion pattern of the Vela OB2 association. 

Notable in the $Z$ vs $Y$ plots is the change in the apparent `tilt' of the Vela OB2 association in the past and in the future, indicating that the association is rotating. This plane is of course the same plane in which we observed a 9$\sigma$ correlation between position ($Y$) and velocity ($W$, the component of velocity in the $Z$ direction) that indicates rotation (Section \ref{rotation}). This provides an important verification of the signature of rotation we observed.

\subsection{Identifying candidate runaway stars}

There are a number of confirmed young stars in our sample whose kinematics are distinct from that of either the Vela OB2 association or the three clusters identified. The PMs of some of these sources point away from the clusters in the region, suggesting that some may be "runaway stars" that have been ejected from one of the clusters due to dynamical interactions. 

To identify runaway stars we consider the past motion of these stars relative to each of the star clusters to see if their traceback intercepts with the half-mass radius of each cluster. We perform this traceback in two dimensions, using a linear projection coordinate frame \citep{helmi18} with a set distance (400 pc). This allowed us to identify 12 sources as candidate runaways whose past motion intercepts with one (or more) of the star clusters.

To verify these runaways we perform three tests to confirm the source as a runaway star. First we require that the SED-fitted age of the star is broadly consistent with the age of the cluster it was ejected from ($\pm 40\%$ of cluster age from Section 3.3). Approximately half of our candidate runaway stars passed this test, with the majority of failures being candidates ejectees from NGC 2547 which were found to be much younger than this cluster. The second test requires that the ejection timescale for the star (the time required for it to transit to its current position) is less than or equal to the age of the cluster. Due to the proximity of these sources to the cluster all of our stars pass the ejection timescale test. The final test requires that the relative line of sight distance between the star and the cluster is consistent with their relative RVs. Most sources were found to fail this test, either because their relative RV was in the wrong direction or was too small or too large to be consistent with being ejected from the cluster.

The only source that passes all of our tests is {\it Gaia} ID 5530691754285644032, which has an SED age ($\sim$17~Myr) and kinematics consistent with having been recently ejected ($0.07 \pm 0.01$ Myr) from the P Puppis cluster at a velocity of $9.43 \pm 0.01$ kms$^{-1}$. This would classify the source more as a 'walk-away' rather than a runaway \citep{demink14,schoettler20}. The remaining sources may be examples of kinematically 'hot' young stars \citep{binks20} or they may be associated with NGC 2547, as previously noted.

\section{Discussion and future work}
\label{section_discussion}

We have studied the 3D dynamics of a group of spectroscopically-confirmed young stars. Our main results are as follows:
\begin{itemize}
	\item We have identified considerable substructure in the form of multiple distinct groups around Vela OB2, namely the known clusters $\gamma$ Vel and NGC 2547, a previously poorly-studied cluster, P Puppis, and the association itself.
	\item We calculate velocity dispersions and virial masses for all these groups, finding significant anisotropy for Vela OB2 and the P Puppis cluster. The small 3D velocity dispersions of the $\gamma$ Vel, NGC 2547 and P Puppis clusters indicate that they are likely gravitationally bound. We calculate virial masses of $272^{+95}_{-80}$, $285^{+190}_{-130}$ and $133^{+75}_{-63}$ M$_{\odot}$ respectively.
	\item We find significant evidence ($>11\sigma$) of expansion for the Vela OB2 association in all directions, though it is somewhat anisotropic. We calculate a kinematic traceback age of 13-24 Myr for Vela OB2 based on its expansion pattern, which is in good agreement with its literature age of 15-20 Myr (Section 3.3).
	\item We have used an epicycle approximation to investigate the relative positions of these groups up to 20 Myrs into the past. We find that NGC 2547 is an interloper to Vela OB2 and is just passing through the region at the present time. Both the $\gamma$ Vel and P Puppis clusters formed within the Vela OB2 association. 
	\item We have identified one likely runaway star whose 3D position, kinematics and age are consistent with being ejected from the P Puppis cluster. All other candidate PM runaways were dismissed based on their age or RV, highlighting the value of using spectroscopy to validate runaway stars.
\end{itemize}

We now discuss the implications of these results in the context of the formation and evolution of OB associations and star clusters.

\subsection{Structure of Vela OB2 and nearby clusters}

Previous studies have not resolved the full 3D structure and dynamics of Vela OB2 and its surrounding clusters. Pre-{\it Gaia} studies of the region hinted at the existence of a sparse, widespread population of young stars \citep{jeffries14,sacco15} but it was not until the availability of {\it Gaia} DR2 proper motions and parallaxes that a structural and kinematic investigation over the whole region could be done \citep{cantatgaudin19b}. Even then, radial velocities were lacking.

We have found that Vela OB2 is not a single homogenous entity. Rather, it is a highly substructured complex containing multiple clusters with a range of ages, surrounded by a widespread and dispersed population of young stars, exhibiting complex dynamics. We have confirmed that the sparse population surrounding the $\gamma$ Vel and NGC 2547 clusters \citep{jeffries14,sacco15} belongs to the Vela OB2 association and have also identified a third cluster in the region, the P Puppis cluster.

By tracing back the motion of stars and clusters we have been able to recreate the spatial distribution of stars in the region at their approximate time of birth. We have found that the NGC 2547 cluster and the `outlier' group are interlopers in the Vela OB2 region, having formed $>$100 pc away and are only transiting the region at the current time.

We have constrained the initial configuration of Vela OB2 and probed its future evolution by tracing backwards and forwards in time the motion of stars in Vela OB2 and the bulk motion of the clusters in its vicinity. This, combined with the older age of NGC 2547 and the outlier group relative to the rest of the Vela region, confirms that they are unrelated to the Vela OB2 association and its clusters.

On the other hand, the $\gamma$ Vel and P Puppis clusters appear to have formed within the volume of the Vela OB2 association, indicating that they originated as compact substructures of Vela OB2. This is supported by their similar ages to Vela OB2. The $\gamma$ Vel cluster appears to have formed relatively centrally within the association, and at a similar time to it, suggesting its formation may have been strongly linked to the formation of the association as a whole, whereas the P Puppis cluster is younger and appears to have formed on the edge of the association, suggesting a total period of star formation up to 10 Myr within the region.

\subsection{Expansion of the association}

We found strong evidence ($>11\sigma$ in all directions) for expansion of the Vela OB2 association that is not apparent in any of the clusters. This is in line with expectations since Vela OB2 was predicted to be an unbound association, while the compact clusters were likely to be gravitationally bound. The expansion of Vela OB2 is in agreement with past studies that have measured the expansion of the association using PMs \citep{cantatgaudin19a,armstrong20}, though this is the first time the expansion has been measured in 3D and to such a significance (11$\sigma$). These expansion rates were then used to calculate a kinematic age for the association (13-24 Myr), which is in good agreement with literature ages for Vela OB2 (15-20 Myr).

The expansion is also notably anisotropic, arguing against a simple explosive radial expansion pattern from the rapid unbinding of a compact group of stars. This is consistent with findings from other OB associations \citep[see discussion in][]{wright20}.

The $\gamma$ Vel and P Puppis clusters show hints of expansion in the $X$ direction, with 4 and 3 $\sigma$ correlations between position and velocity in this axis, respectively. For the former, some contamination from the co-spatial Vela OB2 association may have contributed to this measurement, though our conservative cluster membership selection process should have minimised this. There are a number of physical explanations for the observed expansion. Firstly, the clusters may be gravitationally unbound and expanding. This would be surprising for a cluster as compact as $\gamma$ Vel however, and for both clusters the expansion is strongly anisotropic. In the Galactic Cartesian coordinate system the $X$ is aligned towards the Galactic center, so the expansion in this direction may be due to tidal shearing, though the timescale for this is typically much older than the age of $\gamma$ Vel. It is also possible that the P Puppis and $\gamma$ Vel clusters are being pulled in a certain direction by the mass of the Vela OB2 association or its primordial molecular cloud that correlates strongly with the X direction, though this is strongly dependent on the arrangement of mass in the vicinity of the clusters.

We also observe significant signatures of rotation in the Vela OB2 association in $Y$ vs $W$ and $Z$ vs $V$, which are both in the $Y$-$Z$ plane, but acting in opposite directions. This could be due to subgroups within the association moving in opposite directions and causing the appearance of bimodal rotation. However it could also be caused by the combination of the elongation of the association in the $Y$-$Z$ plane (Figure~\ref{traceback_wide}) and its expansion, which can give signatures of rotation that are actually caused by expansion.
In either case there is certainly some degree of rotation, or residual angular momentum, in the association, as evidenced by the change in inclination of the association over time (see $Y$-$Z$ plots in Fig. \ref{traceback_wide}).

\section{Conclusions}

We have presented the first 6D kinematic study of the Vela OB2 association by combining {\it Gaia} astrometry with spectroscopic RVs for 410 spectroscopically-confirmed young stars. We used this data to study the kinematics of the Vela OB2 association and its constituent clusters.

We separated the sample into multiple kinematic groups, including the $\gamma$ Vel, NGC 2547 and P Puppis clusters, and the Vela OB2 association. We measure 3D velocity dispersions, which for the P Puppis cluster and the Vela OB2 association show evidence of anisotropy, indicating that these groups are dynamically un-evolved, while $\gamma$ Vel and NGC~2547 are consistent with isotropy. We calculate expansion gradients and find very significant evidence for expansion in Vela OB2 in all three dimensions, which is close to isotropy when the component of velocity in the radial direction from the association center is isolated. The $\gamma$ Vel and P Puppis clusters also show some evidence for expansion. We use an epicycle approximation to trace back the positions of stars into the past. We find that the NGC 2547 cluster likely originated $>$100 pc away from Vela OB2 and is an interloper at present, while the $\gamma$ Vel and P Puppis clusters appear to have formed within the Vela OB2 association.

We have established that the Vela OB2 association is highly substructured, both spatially and kinematically. It contains multiple open clusters of different ages with distinct kinematics, which are surrounded by a sparse, expanding population. This raises questions about other associations, do they typically form with such substructure and over what timescales is it erased? Do all OB association contains compact clusters within them, and are all open clusters surrounded by a sparse OB association? The results we present contribute to the growing wealth of evidence that OB associations are not the remnants of initially bound clusters, but instead form across extended regions with considerable substructure over timescales of up to $\sim$10 Myr. 

\section{Acknowledgments}

This work has made use of data from the ESA space mission Gaia (http://www.cosmos.esa.int/gaia), processed by the Gaia Data Processing and Analysis Consortium (DPAC, http://www.cosmos.esa.int/web/gaia/dpac/consortium). Funding for DPAC has been provided by national institutions, in particular the institutions participating in the Gaia Multilateral Agreement. This research made use of the Simbad and Vizier catalogue access tools (provided by CDS, Strasbourg, France), Astropy \citep{astr13} and TOPCAT \citep{tayl05}. 

\section{Data Availability}

The data underlying this article, both the spectroscopic parameters for observed targets and the kinematic properties of Vela OB2, NGC 2547, P Puppis and $\gamma$ cluster, is
available as supplementary material.

\bibliographystyle{mn2e}
\bibliography{references2}

\begin{thebibliography}{85}
\expandafter\ifx\csname natexlab\endcsname\relax\def\natexlab#1{#1}\fi

\bibitem[{{AAO Software Team}(2015)}]{2dfdr}
{AAO Software Team}, 2015, ascl:1505.015

\bibitem[{Ambartsumian(1947)}]{ambartsumian47}
Ambartsumian V., 1947, Acad. Sci. Armenian SSR, Yerevan

\bibitem[{{Andr{\'e}} {et~al}\mbox{.}(2010){Andr{\'e}}, {Men'shchikov},
  {Bontemps}, {K{\"o}nyves}, {Motte}, {Schneider}, {Didelon}, {Minier},
  {Saraceno}, {Ward-Thompson}, {di Francesco}, {White}, {Molinari}, {Testi},
  {Abergel}, {Griffin}, {Henning}, {Royer}, {Mer{\'{\i}}n}, {Vavrek}, {Attard},
  {Arzoumanian}, {Wilson}, {Ade}, {Aussel}, {Baluteau}, {Benedettini},
  {Bernard}, {Blommaert}, {Cambr{\'e}sy}, {Cox}, {di Giorgio}, {Hargrave},
  {Hennemann}, {Huang}, {Kirk}, {Krause}, {Launhardt}, {Leeks}, {Le Pennec},
  {Li}, {Martin}, {Maury}, {Olofsson}, {Omont}, {Peretto}, {Pezzuto}, {Prusti},
  {Roussel}, {Russeil}, {Sauvage}, {Sibthorpe}, {Sicilia-Aguilar}, {Spinoglio},
  {Waelkens}, {Woodcraft}, \& {Zavagno}}]{andre10}
{Andr{\'e}} P. {et~al.}, 2010, \aap, 518, L102

\bibitem[{{Armstrong} {et~al}\mbox{.}(2018){Armstrong}, {Wright}, \&
  {Jeffries}}]{armstrong18}
{Armstrong} J.~J., {Wright} N.~J., {Jeffries} R.~D., 2018, \mnras, 480, L121

\bibitem[{{Armstrong} {et~al}\mbox{.}(2020){Armstrong}, {Wright}, {Jeffries},
  \& {Jackson}}]{armstrong20}
{Armstrong} J.~J., {Wright} N.~J., {Jeffries} R.~D., {Jackson} R.~J., 2020,
  \mnras, 494, 4794

\bibitem[{{Astropy Collaboration} {et~al}\mbox{.}(2013){Astropy Collaboration},
  {Robitaille}, {Tollerud}, {Greenfield}, {Droettboom}, {Bray}, {Aldcroft},
  {Davis}, {Ginsburg}, {Price-Whelan}, {Kerzendorf}, {Conley}, {Crighton},
  {Barbary}, {Muna}, {Ferguson}, {Grollier}, {Parikh}, {Nair}, {Unther},
  {Deil}, {Woillez}, {Conseil}, {Kramer}, {Turner}, {Singer}, {Fox}, {Weaver},
  {Zabalza}, {Edwards}, {Azalee Bostroem}, {Burke}, {Casey}, {Crawford},
  {Dencheva}, {Ely}, {Jenness}, {Labrie}, {Lim}, {Pierfederici}, {Pontzen},
  {Ptak}, {Refsdal}, {Servillat}, \& {Streicher}}]{astr13}
{Astropy Collaboration} {et~al.}, 2013, \aap, 558, A33

\bibitem[{{Bailer-Jones} {et~al}\mbox{.}(2018){Bailer-Jones}, {Rybizki},
  {Fouesneau}, {Mantelet}, \& {Andrae}}]{bailerjones18}
{Bailer-Jones} C.~A.~L., {Rybizki} J., {Fouesneau} M., {Mantelet} G., {Andrae}
  R., 2018, \aj, 156, 58

\bibitem[{{Baraffe} {et~al}\mbox{.}(2015){Baraffe}, {Homeier}, {Allard}, \&
  {Chabrier}}]{baraffe15}
{Baraffe} I., {Homeier} D., {Allard} F., {Chabrier} G., 2015, \aap, 577, A42

\bibitem[{{Beccari} {et~al}\mbox{.}(2018){Beccari}, {Boffin}, {Jerabkova},
  {Wright}, {Kalari}, {Carraro}, {De Marchi}, \& {de Wit}}]{beccari18}
{Beccari} G., {Boffin} H. M.~J., {Jerabkova} T., {Wright} N.~J., {Kalari}
  V.~M., {Carraro} G., {De Marchi} G., {de Wit} W.-J., 2018, \mnras, 481, L11

\bibitem[{{Binks} {et~al}\mbox{.}(2020){Binks}, {Jeffries}, \&
  {Wright}}]{binks20}
{Binks} A.~S., {Jeffries} R.~D., {Wright} N.~J., 2020, \mnras, 494, 2429

\bibitem[{{Blaauw}(1964)}]{blaauw64}
{Blaauw} A., 1964, \araa, 2, 213

\bibitem[{{Brice{\~n}o} {et~al}\mbox{.}(2007){Brice{\~n}o}, {Hartmann},
  {Hern{\'a}ndez}, {Calvet}, {Vivas}, {Furesz}, \& {Szentgyorgyi}}]{briceno07}
{Brice{\~n}o} C., {Hartmann} L., {Hern{\'a}ndez} J., {Calvet} N., {Vivas}
  A.~K., {Furesz} G., {Szentgyorgyi} A., 2007, \apj, 661, 1119

\bibitem[{{Caballero} \& {Dinis}(2008)}]{caballero08}
{Caballero} J.~A., {Dinis} L., 2008, Astronomische Nachrichten, 329, 801

\bibitem[{{Cantat-Gaudin} {et~al}\mbox{.}(2019{\natexlab{a}}){Cantat-Gaudin},
  {Jordi}, {Wright}, {Armstrong}, {Vallenari}, {Balaguer-N{\'u}{\~n}ez},
  {Ramos}, {Bossini}, {Padoan}, {Pelkonen}, {Mapelli}, \&
  {Jeffries}}]{cantatgaudin19b}
{Cantat-Gaudin} T. {et~al.}, 2019{\natexlab{a}}, \aap, 626, A17

\bibitem[{{Cantat-Gaudin} {et~al}\mbox{.}(2019{\natexlab{b}}){Cantat-Gaudin},
  {Mapelli}, {Balaguer-N{\'u}{\~n}ez}, {Jordi}, {Sacco}, \&
  {Vallenari}}]{cantatgaudin19a}
{Cantat-Gaudin} T., {Mapelli} M., {Balaguer-N{\'u}{\~n}ez} L., {Jordi} C.,
  {Sacco} G., {Vallenari} A., 2019{\natexlab{b}}, \aap, 621, A115

\bibitem[{{Chen} {et~al}\mbox{.}(2014){Chen}, {Girardi}, {Bressan}, {Marigo},
  {Barbieri}, \& {Kong}}]{chen14}
{Chen} Y., {Girardi} L., {Bressan} A., {Marigo} P., {Barbieri} M., {Kong} X.,
  2014, \mnras, 444, 2525

\bibitem[{{Damiani} {et~al}\mbox{.}(2017){Damiani}, {Prisinzano}, {Jeffries},
  {Sacco}, {Randich}, \& {Micela}}]{damiani17}
{Damiani} F., {Prisinzano} L., {Jeffries} R.~D., {Sacco} G.~G., {Randich} S.,
  {Micela} G., 2017, \aap, 602, L1

\bibitem[{{Damiani} {et~al}\mbox{.}(2014){Damiani}, {Prisinzano}, {Micela},
  {Randich}, {Gilmore}, {Drew}, {Jeffries}, {Fr{\'e}mat}, {Alfaro}, {Bensby},
  {Bragaglia}, {Flaccomio}, {Lanzafame}, {Pancino}, {Recio-Blanco}, {Sacco},
  {Smiljanic}, {Jackson}, {de Laverny}, {Morbidelli}, {Worley}, {Hourihane},
  {Costado}, {Jofr{\'e}}, {Lind}, \& {Maiorca}}]{damiani14}
{Damiani} F. {et~al.}, 2014, \aap, 566, A50

\bibitem[{{Damiani} {et~al}\mbox{.}(2019){Damiani}, {Prisinzano}, {Pillitteri},
  {Micela}, \& {Sciortino}}]{damiani19}
{Damiani} F., {Prisinzano} L., {Pillitteri} I., {Micela} G., {Sciortino} S.,
  2019, \aap, 623, A112

\bibitem[{{de Mink} {et~al}\mbox{.}(2014){de Mink}, {Sana}, {Langer}, {Izzard},
  \& {Schneider}}]{demink14}
{de Mink} S.~E., {Sana} H., {Langer} N., {Izzard} R.~G., {Schneider} F.~R.~N.,
  2014, \apj, 782, 7

\bibitem[{{de Zeeuw} {et~al}\mbox{.}(1999){de Zeeuw}, {Hoogerwerf}, {de
  Bruijne}, {Brown}, \& {Blaauw}}]{dezeeuw99}
{de Zeeuw} P.~T., {Hoogerwerf} R., {de Bruijne} J.~H.~J., {Brown} A.~G.~A.,
  {Blaauw} A., 1999, \aj, 117, 354

\bibitem[{{El-Badry} {et~al}\mbox{.}(2021){El-Badry}, {Rix}, \&
  {Heintz}}]{elbadry21}
{El-Badry} K., {Rix} H.-W., {Heintz} T.~M., 2021, \mnras, 506, 2269

\bibitem[{Ester {et~al}\mbox{.}(1996)Ester, Kriegel, Sander, Xu,
  {et~al.}}]{ester96}
Ester M., Kriegel H.-P., Sander J., Xu X., {et~al.}, 1996, in Kdd, Vol.~96, pp.
  226--231

\bibitem[{{Feast} \& {Whitelock}(1997)}]{feast97}
{Feast} M., {Whitelock} P., 1997, \mnras, 291, 683

\bibitem[{Foreman-Mackey {et~al}\mbox{.}(2013)Foreman-Mackey, Hogg, Lang, \&
  Goodman}]{emcee}
Foreman-Mackey D., Hogg D.~W., Lang D., Goodman J., 2013, Publications of the
  Astronomical Society of the Pacific, 125, 306

\bibitem[{{Franciosini} {et~al}\mbox{.}(2018){Franciosini}, {Sacco},
  {Jeffries}, {Damiani}, {Roccatagliata}, {Fedele}, \&
  {Randich}}]{franciosini18}
{Franciosini} E., {Sacco} G.~G., {Jeffries} R.~D., {Damiani} F.,
  {Roccatagliata} V., {Fedele} D., {Randich} S., 2018, \aap, 616, L12

\bibitem[{{Fuchs} {et~al}\mbox{.}(2006){Fuchs}, {Breitschwerdt}, {de Avillez},
  {Dettbarn}, \& {Flynn}}]{fuchs06}
{Fuchs} B., {Breitschwerdt} D., {de Avillez} M.~A., {Dettbarn} C., {Flynn} C.,
  2006, \mnras, 373, 993

\bibitem[{{Gaia Collaboration} {et~al}\mbox{.}(2018){Gaia Collaboration},
  {Brown}, {Vallenari}, {Prusti}, {de Bruijne}, {Babusiaux}, {Bailer-Jones},
  {Biermann}, {Evans}, \& {Eyer}}]{gaiadr2summary}
{Gaia Collaboration} {et~al.}, 2018, \aap, 616, A1

\bibitem[{{Gaia Collaboration} {et~al}\mbox{.}(2021){Gaia Collaboration},
  {Brown}, {Vallenari}, {Prusti}, {de Bruijne}, {Babusiaux}, {Biermann},
  {Creevey}, {Evans}, {Eyer}, {Hutton}, {Jansen}, {Jordi}, {Klioner},
  {Lammers}, {Lindegren}, {Luri}, {Mignard}, {Panem}, {Pourbaix}, {Randich},
  {Sartoretti}, {Soubiran}, {Walton}, {Arenou}, {Bailer-Jones}, {Bastian},
  {Cropper}, {Drimmel}, {Katz}, {Lattanzi}, {van Leeuwen}, {Bakker},
  {Cacciari}, {Casta{\~n}eda}, {De Angeli}, {Ducourant}, {Fabricius},
  {Fouesneau}, {Fr{\'e}mat}, {Guerra}, {Guerrier}, {Guiraud}, {Jean-Antoine
  Piccolo}, {Masana}, {Messineo}, {Mowlavi}, {Nicolas}, {Nienartowicz},
  {Pailler}, {Panuzzo}, {Riclet}, {Roux}, {Seabroke}, {Sordo}, {Tanga},
  {Th{\'e}venin}, {Gracia-Abril}, {Portell}, {Teyssier}, {Altmann}, {Andrae},
  {Bellas-Velidis}, {Benson}, {Berthier}, {Blomme}, {Brugaletta}, {Burgess},
  {Busso}, {Carry}, {Cellino}, {Cheek}, {Clementini}, {Damerdji}, {Davidson},
  {Delchambre}, {Dell'Oro}, {Fern{\'a}ndez-Hern{\'a}ndez}, {Galluccio},
  {Garc{\'\i}a-Lario}, {Garcia-Reinaldos}, {Gonz{\'a}lez-N{\'u}{\~n}ez},
  {Gosset}, {Haigron}, {Halbwachs}, {Hambly}, {Harrison}, {Hatzidimitriou},
  {Heiter}, {Hern{\'a}ndez}, {Hestroffer}, {Hodgkin}, {Holl}, {Jan{\ss}en},
  {Jevardat de Fombelle}, {Jordan}, {Krone-Martins}, {Lanzafame},
  {L{\"o}ffler}, {Lorca}, {Manteiga}, {Marchal}, {Marrese}, {Moitinho}, {Mora},
  {Muinonen}, {Osborne}, {Pancino}, {Pauwels}, {Petit}, {Recio-Blanco},
  {Richards}, {Riello}, {Rimoldini}, {Robin}, {Roegiers}, {Rybizki}, {Sarro},
  {Siopis}, {Smith}, {Sozzetti}, {Ulla}, {Utrilla}, {van Leeuwen}, {van
  Reeven}, {Abbas}, {Abreu Aramburu}, {Accart}, {Aerts}, {Aguado}, {Ajaj},
  {Altavilla}, {{\'A}lvarez}, {{\'A}lvarez Cid-Fuentes}, {Alves}, {Anderson},
  {Anglada Varela}, {Antoja}, {Audard}, {Baines}, {Baker},
  {Balaguer-N{\'u}{\~n}ez}, {Balbinot}, {Balog}, {Barache}, {Barbato},
  {Barros}, {Barstow}, {Bartolom{\'e}}, {Bassilana}, {Bauchet},
  {Baudesson-Stella}, {Becciani}, {Bellazzini}, {Bernet}, {Bertone}, {Bianchi},
  {Blanco-Cuaresma}, {Boch}, {Bombrun}, {Bossini}, {Bouquillon}, {Bragaglia},
  {Bramante}, {Breedt}, {Bressan}, {Brouillet}, {Bucciarelli}, {Burlacu},
  {Busonero}, {Butkevich}, {Buzzi}, {Caffau}, {Cancelliere}, {C{\'a}novas},
  {Cantat-Gaudin}, {Carballo}, {Carlucci}, {Carnerero}, {Carrasco},
  {Casamiquela}, {Castellani}, {Castro-Ginard}, {Castro Sampol}, {Chaoul},
  {Charlot}, {Chemin}, {Chiavassa}, {Cioni}, {Comoretto}, {Cooper}, {Cornez},
  {Cowell}, {Crifo}, {Crosta}, {Crowley}, {Dafonte}, {Dapergolas}, {David},
  {David}, {de Laverny}, {De Luise}, {De March}, {De Ridder}, {de Souza}, {de
  Teodoro}, {de Torres}, {del Peloso}, {del Pozo}, {Delbo}, {Delgado},
  {Delgado}, {Delisle}, {Di Matteo}, {Diakite}, {Diener}, {Distefano},
  {Dolding}, {Eappachen}, {Edvardsson}, {Enke}, {Esquej}, {Fabre}, {Fabrizio},
  {Faigler}, {Fedorets}, {Fernique}, {Fienga}, {Figueras}, {Fouron},
  {Fragkoudi}, {Fraile}, {Franke}, {Gai}, {Garabato}, {Garcia-Gutierrez},
  {Garc{\'\i}a-Torres}, {Garofalo}, {Gavras}, {Gerlach}, {Geyer}, {Giacobbe},
  {Gilmore}, {Girona}, {Giuffrida}, {Gomel}, {Gomez}, {Gonzalez-Santamaria},
  {Gonz{\'a}lez-Vidal}, {Granvik}, {Guti{\'e}rrez-S{\'a}nchez}, {Guy},
  {Hauser}, {Haywood}, {Helmi}, {Hidalgo}, {Hilger}, {H{\l}adczuk}, {Hobbs},
  {Holland}, {Huckle}, {Jasniewicz}, {Jonker}, {Juaristi Campillo}, {Julbe},
  {Karbevska}, {Kervella}, {Khanna}, {Kochoska}, {Kontizas}, {Kordopatis},
  {Korn}, {Kostrzewa-Rutkowska}, {Kruszy{\'n}ska}, {Lambert}, {Lanza}, {Lasne},
  {Le Campion}, {Le Fustec}, {Lebreton}, {Lebzelter}, {Leccia}, {Leclerc},
  {Lecoeur-Taibi}, {Liao}, {Licata}, {Lindstr{\o}m}, {Lister}, {Livanou},
  {Lobel}, {Madrero Pardo}, {Managau}, {Mann}, {Marchant}, {Marconi}, {Marcos
  Santos}, {Marinoni}, {Marocco}, {Marshall}, {Martin Polo},
  {Mart{\'\i}n-Fleitas}, {Masip}, {Massari}, {Mastrobuono-Battisti}, {Mazeh},
  {McMillan}, {Messina}, {Michalik}, {Millar}, {Mints}, {Molina}, {Molinaro},
  {Moln{\'a}r}, {Montegriffo}, {Mor}, {Morbidelli}, {Morel}, {Morris},
  {Mulone}, {Munoz}, {Muraveva}, {Murphy}, {Musella}, {Noval}, {Ord{\'e}novic},
  {Orr{\`u}}, {Osinde}, {Pagani}, {Pagano}, {Palaversa}, {Palicio}, {Panahi},
  {Pawlak}, {Pe{\~n}alosa Esteller}, {Penttil{\"a}}, {Piersimoni}, {Pineau},
  {Plachy}, {Plum}, {Poggio}, {Poretti}, {Poujoulet}, {Pr{\v{s}}a}, {Pulone},
  {Racero}, {Ragaini}, {Rainer}, {Raiteri}, {Rambaux}, {Ramos}, {Ramos-Lerate},
  {Re Fiorentin}, {Regibo}, {Reyl{\'e}}, {Ripepi}, {Riva}, {Rixon}, {Robichon},
  {Robin}, {Roelens}, {Rohrbasser}, {Romero-G{\'o}mez}, {Rowell}, {Royer},
  {Rybicki}, {Sadowski}, {Sagrist{\`a} Sell{\'e}s}, {Sahlmann}, {Salgado},
  {Salguero}, {Samaras}, {Sanchez Gimenez}, {Sanna}, {Santove{\~n}a},
  {Sarasso}, {Schultheis}, {Sciacca}, {Segol}, {Segovia}, {S{\'e}gransan},
  {Semeux}, {Shahaf}, {Siddiqui}, {Siebert}, {Siltala}, {Slezak}, {Smart},
  {Solano}, {Solitro}, {Souami}, {Souchay}, {Spagna}, {Spoto}, {Steele},
  {Steidelm{\"u}ller}, {Stephenson}, {S{\"u}veges}, {Szabados}, {Szegedi-Elek},
  {Taris}, {Tauran}, {Taylor}, {Teixeira}, {Thuillot}, {Tonello}, {Torra},
  {Torra}, {Turon}, {Unger}, {Vaillant}, {van Dillen}, {Vanel}, {Vecchiato},
  {Viala}, {Vicente}, {Voutsinas}, {Weiler}, {Wevers}, {Wyrzykowski}, {Yoldas},
  {Yvard}, {Zhao}, {Zorec}, {Zucker}, {Zurbach}, \& {Zwitter}}]{Gaiaedr3}
{Gaia Collaboration} {et~al.}, 2021, \aap, 649, A1

\bibitem[{{Gaia Collaboration} {et~al}\mbox{.}(2016){Gaia Collaboration},
  {Prusti}, {de Bruijne}, {Brown}, {Vallenari}, {Babusiaux}, {Bailer-Jones},
  {Bastian}, {Biermann}, \& {Evans}}]{gaia16}
{Gaia Collaboration} {et~al.}, 2016, \aap, 595, A1

\bibitem[{{Gilmore} {et~al}\mbox{.}(2012){Gilmore}, {Randich}, {Asplund},
  {Binney}, {Bonifacio}, {Drew}, {Feltzing}, {Ferguson}, {Jeffries}, {Micela},
  \& et~al.}]{gilmore12}
{Gilmore} G. {et~al.}, 2012, The Messenger, 147, 25

\bibitem[{{Guti{\'e}rrez} \& {Ramos-Chernenko}(2022)}]{gutierrez22}
{Guti{\'e}rrez} C.~M., {Ramos-Chernenko} N., 2022, \apj, 929, 29

\bibitem[{{Helmi} {et~al}\mbox{.}(2018){Helmi}, {Babusiaux}, {Koppelman},
  {Massari}, {Veljanoski}, \& {Brown}}]{helmi18}
{Helmi} A., {Babusiaux} C., {Koppelman} H.~H., {Massari} D., {Veljanoski} J.,
  {Brown} A. G.~A., 2018, \nat, 563, 85

\bibitem[{{Hills}(1980)}]{hills80}
{Hills} J.~G., 1980, \apj, 235, 986

\bibitem[{{Hogg} {et~al}\mbox{.}(2010){Hogg}, {Bovy}, \& {Lang}}]{hogg10}
{Hogg} D.~W., {Bovy} J., {Lang} D., 2010, arXiv e-prints, arXiv:1008.4686

\bibitem[{{Holmberg} \& {Flynn}(2004)}]{holmberg04}
{Holmberg} J., {Flynn} C., 2004, \mnras, 352, 440

\bibitem[{{Jackson} {et~al}\mbox{.}(2018){Jackson}, {Deliyannis}, \&
  {Jeffries}}]{jackson18}
{Jackson} R.~J., {Deliyannis} C.~P., {Jeffries} R.~D., 2018, \mnras, 476, 3245

\bibitem[{{Jackson} {et~al}\mbox{.}(2022){Jackson}, {Jeffries}, {Wright},
  {Randich}, {Sacco}, {Bragaglia}, {Hourihane}, {Tognelli}, {Degl'Innocenti},
  {Prada Moroni}, {Gilmore}, {Bensby}, {Pancino}, {Smiljanic}, {Bergemann},
  {Carraro}, {Franciosini}, {Gonneau}, {Jofr{\'e}}, {Lewis}, {Magrini},
  {Morbidelli}, {Prisinzano}, {Worley}, {Zaggia}, {Tautvai{\v{s}}iene},
  {Guti{\'e}rrez Albarr{\'a}n}, {Montes}, \& {Jim{\'e}nez-Esteban}}]{jackson22}
{Jackson} R.~J. {et~al.}, 2022, \mnras, 509, 1664

\bibitem[{{Jeffries} {et~al}\mbox{.}(2014){Jeffries}, {Jackson}, {Cottaar},
  {Koposov}, {Lanzafame}, {Meyer}, {Prisinzano}, {Randich}, {Sacco},
  {Brugaletta}, {Caramazza}, {Damiani}, {Franciosini}, {Frasca}, {Gilmore},
  {Feltzing}, {Micela}, {Alfaro}, {Bensby}, {Pancino}, {Recio-Blanco}, {de
  Laverny}, {Lewis}, {Magrini}, {Morbidelli}, {Costado}, {Jofr{\'e}},
  {Klutsch}, {Lind}, \& {Maiorca}}]{jeffries14}
{Jeffries} R.~D. {et~al.}, 2014, \aap, 563, A94

\bibitem[{{Jeffries} {et~al}\mbox{.}(2017){Jeffries}, {Jackson}, {Franciosini},
  {Rand ich}, {Barrado}, {Frasca}, {Klutsch}, {Lanzafame}, {Prisinzano},
  {Sacco}, {Gilmore}, {Vallenari}, {Alfaro}, {Koposov}, {Pancino}, {Bayo},
  {Casey}, {Costado}, {Damiani}, {Hourihane}, {Lewis}, {Jofre}, {Magrini},
  {Monaco}, {Morbidelli}, {Worley}, {Zaggia}, \& {Zwitter}}]{jeffries17}
{Jeffries} R.~D. {et~al.}, 2017, \mnras, 464, 1456

\bibitem[{{Jeffries} {et~al}\mbox{.}(2021){Jeffries}, {Jackson}, {Sun}, \&
  {Deliyannis}}]{jeffries21}
{Jeffries} R.~D., {Jackson} R.~J., {Sun} Q., {Deliyannis} C.~P., 2021, \mnras,
  500, 1158

\bibitem[{{Jeffries} {et~al}\mbox{.}(2004){Jeffries}, {Naylor}, {Devey}, \&
  {Totten}}]{jeffries04}
{Jeffries} R.~D., {Naylor} T., {Devey} C.~R., {Totten} E.~J., 2004, \mnras,
  351, 1401

\bibitem[{{Jeffries} {et~al}\mbox{.}(2009){Jeffries}, {Naylor}, {Walter},
  {Pozzo}, \& {Devey}}]{jeffries09}
{Jeffries} R.~D., {Naylor} T., {Walter} F.~M., {Pozzo} M.~P., {Devey} C.~R.,
  2009, \mnras, 393, 538

\bibitem[{{Jeffries} \& {Oliveira}(2005)}]{jeffries05}
{Jeffries} R.~D., {Oliveira} J.~M., 2005, \mnras, 358, 13

\bibitem[{{Johnson} \& {Soderblom}(1987)}]{johnson87}
{Johnson} D.~R.~H., {Soderblom} D.~R., 1987, \aj, 93, 864

\bibitem[{{Karim} \& {Mamajek}(2017)}]{karim17}
{Karim} M.~T., {Mamajek} E.~E., 2017, \mnras, 465, 472

\bibitem[{{Kroupa} {et~al}\mbox{.}(2001){Kroupa}, {Aarseth}, \&
  {Hurley}}]{kroupa01b}
{Kroupa} P., {Aarseth} S., {Hurley} J., 2001, \mnras, 321, 699

\bibitem[{{Kruijssen} {et~al}\mbox{.}(2012){Kruijssen}, {Maschberger},
  {Moeckel}, {Clarke}, {Bastian}, \& {Bonnell}}]{kruijssen12}
{Kruijssen} J.~M.~D., {Maschberger} T., {Moeckel} N., {Clarke} C.~J., {Bastian}
  N., {Bonnell} I.~A., 2012, \mnras, 419, 841

\bibitem[{{Kurucz}(1992)}]{Kurucz1992a}
{Kurucz} R.~L., 1992, in IAU Symposium, Vol. 149, The Stellar Populations of
  Galaxies, {Barbuy} B., {Renzini} A., eds., p. 225

\bibitem[{{Lewis} {et~al}\mbox{.}(2002){Lewis}, {Cannon}, {Taylor},
  {Glazebrook}, {Bailey}, {Baldry}, {Barton}, {Bridges}, {Dalton}, {Farrell},
  {Gray}, {Lankshear}, {McCowage}, {Parry}, {Sharples}, {Shortridge}, {Smith},
  {Stevenson}, {Straede}, {Waller}, {Whittard}, {Wilcox}, \& {Willis}}]{2dF}
{Lewis} I.~J. {et~al.}, 2002, \mnras, 333, 279

\bibitem[{{Lindegren} {et~al}\mbox{.}(2018){Lindegren}, {Hern{\'a}ndez},
  {Bombrun}, {Klioner}, {Bastian}, {Ramos-Lerate}, {de Torres},
  {Steidelm{\"u}ller}, {Stephenson}, {Hobbs}, {Lammers}, {Biermann}, {Geyer},
  {Hilger}, {Michalik}, {Stampa}, {McMillan}, {Casta{\~n}eda}, {Clotet},
  {Comoretto}, {Davidson}, {Fabricius}, {Gracia}, {Hambly}, {Hutton}, {Mora},
  {Portell}, {van Leeuwen}, {Abbas}, {Abreu}, {Altmann}, {Andrei}, {Anglada},
  {Balaguer-N{\'u}{\~n}ez}, {Barache}, {Becciani}, {Bertone}, {Bianchi},
  {Bouquillon}, {Bourda}, {Br{\"u}semeister}, {Bucciarelli}, {Busonero},
  {Buzzi}, {Cancelliere}, {Carlucci}, {Charlot}, {Cheek}, {Crosta}, {Crowley},
  {de Bruijne}, {de Felice}, {Drimmel}, {Esquej}, {Fienga}, {Fraile}, {Gai},
  {Garralda}, {Gonz{\'a}lez-Vidal}, {Guerra}, {Hauser}, {Hofmann}, {Holl},
  {Jordan}, {Lattanzi}, {Lenhardt}, {Liao}, {Licata}, {Lister}, {L{\"o}ffler},
  {Marchant}, {Martin-Fleitas}, {Messineo}, {Mignard}, {Morbidelli}, {Poggio},
  {Riva}, {Rowell}, {Salguero}, {Sarasso}, {Sciacca}, {Siddiqui}, {Smart},
  {Spagna}, {Steele}, {Taris}, {Torra}, {van Elteren}, {van Reeven}, \&
  {Vecchiato}}]{lindegren18}
{Lindegren} L. {et~al.}, 2018, \aap, 616, A2

\bibitem[{{Littlefair} {et~al}\mbox{.}(2003){Littlefair}, {Naylor}, {Jeffries},
  {Devey}, \& {Vine}}]{littlefair03}
{Littlefair} S.~P., {Naylor} T., {Jeffries} R.~D., {Devey} C.~R., {Vine} S.,
  2003, \mnras, 345, 1205

\bibitem[{M.~Cutri {et~al}\mbox{.}(2003)M.~Cutri, F.~Skrutskie, van Dyk,
  A.~Beichman, M.~Carpenter, Chester, Cambresy, Evans, Fowler, Gizis, Howard,
  \& Huchra}]{cutri03}
M.~Cutri R. {et~al.}, 2003

\bibitem[{{Marigo} {et~al}\mbox{.}(2017){Marigo}, {Girardi}, {Bressan},
  {Rosenfield}, {Aringer}, {Chen}, {Dussin}, {Nanni}, {Pastorelli},
  {Rodrigues}, {Trabucchi}, {Bladh}, {Dalcanton}, {Groenewegen},
  {Montalb{\'a}n}, \& {Wood}}]{marigo17}
{Marigo} P. {et~al.}, 2017, \apj, 835, 77

\bibitem[{{Maschberger}(2013)}]{maschberger13}
{Maschberger} T., 2013, \mnras, 429, 1725

\bibitem[{{McMahon} {et~al}\mbox{.}(2013){McMahon}, {Banerji}, {Gonzalez},
  {Koposov}, {Bejar}, {Lodieu}, {Rebolo}, \& {VHS Collaboration}}]{VHS}
{McMahon} R.~G., {Banerji} M., {Gonzalez} E., {Koposov} S.~E., {Bejar} V.~J.,
  {Lodieu} N., {Rebolo} R., {VHS Collaboration}, 2013, The Messenger, 154, 35

\bibitem[{{Miville-Desch{\^e}nes} \& {Lagache}(2005)}]{IRIS}
{Miville-Desch{\^e}nes} M.-A., {Lagache} G., 2005, \apjs, 157, 302

\bibitem[{{Nikoghosyan} \& {Azatyan}(2019)}]{nikoghosyan19}
{Nikoghosyan} E.~H., {Azatyan} N.~M., 2019, \pasa, 36, e039

\bibitem[{{North} {et~al}\mbox{.}(2007){North}, {Tuthill}, {Tango}, \&
  {Davis}}]{north07}
{North} J.~R., {Tuthill} P.~G., {Tango} W.~J., {Davis} J., 2007, \mnras, 377,
  415

\bibitem[{{Palla} {et~al}\mbox{.}(2005){Palla}, {Randich}, {Flaccomio}, \&
  {Pallavicini}}]{palla05}
{Palla} F., {Randich} S., {Flaccomio} E., {Pallavicini} R., 2005, \apjl, 626,
  L49

\bibitem[{{Pang} {et~al}\mbox{.}(2021){Pang}, {Yu}, {Tang}, {Hong}, {Yuan},
  {Pasquato}, \& {Kouwenhoven}}]{pang21}
{Pang} X., {Yu} Z., {Tang} S.-Y., {Hong} J., {Yuan} Z., {Pasquato} M.,
  {Kouwenhoven} M.~B.~N., 2021, \apj, 923, 20

\bibitem[{{Parker} {et~al}\mbox{.}(2009){Parker}, {Goodwin}, {Kroupa}, \&
  {Kouwenhoven}}]{parker09}
{Parker} R.~J., {Goodwin} S.~P., {Kroupa} P., {Kouwenhoven} M.~B.~N., 2009,
  \mnras, 397, 1577

\bibitem[{{Pozzo} {et~al}\mbox{.}(2000){Pozzo}, {Jeffries}, {Naylor}, {Totten},
  {Harmer}, \& {Kenyon}}]{pozzo00}
{Pozzo} M., {Jeffries} R.~D., {Naylor} T., {Totten} E.~J., {Harmer} S.,
  {Kenyon} M., 2000, \mnras, 313, L23

\bibitem[{{Preibisch} {et~al}\mbox{.}(2002){Preibisch}, {Brown}, {Bridges},
  {Guenther}, \& {Zinnecker}}]{preibisch02}
{Preibisch} T., {Brown} A. G.~A., {Bridges} T., {Guenther} E., {Zinnecker} H.,
  2002, \aj, 124, 404

\bibitem[{{Raghavan} {et~al}\mbox{.}(2010){Raghavan}, {McAlister}, {Henry},
  {Latham}, {Marcy}, {Mason}, {Gies}, {White}, \& {ten
  Brummelaar}}]{raghavan10}
{Raghavan} D. {et~al.}, 2010, \apjs, 190, 1

\bibitem[{{Sacco} {et~al}\mbox{.}(2015){Sacco}, {Jeffries}, {Randich},
  {Franciosini}, {Jackson}, {Cottaar}, {Spina}, {Palla}, {Mapelli}, {Alfaro},
  {Bonito}, {Damiani}, {Frasca}, {Klutsch}, {Lanzafame}, {Bayo}, {Barrado},
  {Jim{\'e}nez-Esteban}, {Gilmore}, {Micela}, {Vallenari}, {Allende Prieto},
  {Flaccomio}, {Carraro}, {Costado}, {Jofr{\'e}}, {Lardo}, {Magrini},
  {Morbidelli}, {Prisinzano}, \& {Sbordone}}]{sacco15}
{Sacco} G.~G. {et~al.}, 2015, \aap, 574, L7

\bibitem[{{Sahu}(1992)}]{sahu92}
{Sahu} M.~S., 1992, PhD thesis, Kapteyn Institute, Postbus 800 9700 AV
  Groningen, The Netherlands

\bibitem[{{S{\'a}nchez} \& {Alfaro}(2009)}]{sanchez09}
{S{\'a}nchez} N., {Alfaro} E.~J., 2009, \apj, 696, 2086

\bibitem[{{Schoettler} {et~al}\mbox{.}(2020){Schoettler}, {de Bruijne},
  {Vaher}, \& {Parker}}]{schoettler20}
{Schoettler} C., {de Bruijne} J., {Vaher} E., {Parker} R.~J., 2020, \mnras,
  495, 3104

\bibitem[{{Sch{\"o}nrich} {et~al}\mbox{.}(2010){Sch{\"o}nrich}, {Binney}, \&
  {Dehnen}}]{schonrich10}
{Sch{\"o}nrich} R., {Binney} J., {Dehnen} W., 2010, \mnras, 403, 1829

\bibitem[{{Sheinis} {et~al}\mbox{.}(2015){Sheinis}, {Anguiano}, {Asplund},
  {Bacigalupo}, {Barden}, {Birchall}, {Bland -Hawthorn}, {Brzeski}, {Cannon},
  {Carollo}, {Case}, {Casey}, {Churilov}, {Warrick}, {Dean}, {De Silva},
  {D'Orazi}, {Duong}, {Farrell}, {Fiegert}, {Freeman}, {Gabriella}, {Gers},
  {Goodwin}, {Gray}, {Green}, {Heald}, {Heijmans}, {Ireland}, {Jones}, {Kafle},
  {Keller}, {Klauser}, {Kondrat}, {Kos}, {Lawrence}, {Lee}, {Mali}, {Martell},
  {Mathews}, {Mayfield}, {Miziarski}, {Muller}, {Pai}, {Patterson}, {Penny},
  {Orr}, {Schlesinger}, {Sharma}, {Shortridge}, {Simpson}, {Smedley}, {Smith},
  {Stafford}, {Staszak}, {Vuong}, {Waller}, {de Boer}, {Xavier}, {Zheng},
  {Zhelem}, {Zucker}, \& {Zwitter}}]{HERMES}
{Sheinis} A. {et~al.}, 2015, Journal of Astronomical Telescopes, Instruments,
  and Systems, 1, 035002

\bibitem[{{Sneden} {et~al}\mbox{.}(2012){Sneden}, {Bean}, {Ivans}, {Lucatello},
  \& {Sobeck}}]{Sneden2012a}
{Sneden} C., {Bean} J., {Ivans} I., {Lucatello} S., {Sobeck} J., 2012,
  ascl:1202.009

\bibitem[{{Soderblom}(2010)}]{soderblom10}
{Soderblom} D.~R., 2010, araa, 48, 581

\bibitem[{{Spina} {et~al}\mbox{.}(2017){Spina}, {Randich}, {Magrini},
  {Jeffries}, {Friel}, {Sacco}, {Pancino}, {Bonito}, {Bravi}, {Franciosini},
  {Klutsch}, {Montes}, {Gilmore}, {Vallenari}, {Bensby}, {Bragaglia},
  {Flaccomio}, {Koposov}, {Korn}, {Lanzafame}, {Smiljanic}, {Bayo}, {Carraro},
  {Casey}, {Costado}, {Damiani}, {Donati}, {Frasca}, {Hourihane}, {Jofr{\'e}},
  {Lewis}, {Lind}, {Monaco}, {Morbidelli}, {Prisinzano}, {Sousa}, {Worley}, \&
  {Zaggia}}]{spina17}
{Spina} L. {et~al.}, 2017, \aap, 601, A70

\bibitem[{{Taylor}(2005)}]{tayl05}
{Taylor} M.~B., 2005, in Astronomical Society of the Pacific Conference Series,
  Vol. 347, Astronomical Data Analysis Software and Systems XIV, {Shopbell} P.,
  {Britton} M., {Ebert} R., eds., p.~29

\bibitem[{{Tobin} {et~al}\mbox{.}(2009){Tobin}, {Hartmann}, {Furesz}, {Mateo},
  \& {Megeath}}]{tobin2009}
{Tobin} J.~J., {Hartmann} L., {Furesz} G., {Mateo} M., {Megeath} S.~T., 2009,
  \apj, 697, 1103

\bibitem[{{Tutukov}(1978)}]{tutukov78}
{Tutukov} A.~V., 1978, \aap, 70, 57

\bibitem[{{Ward} \& {Kruijssen}(2018)}]{ward18}
{Ward} J.~L., {Kruijssen} J.~M.~D., 2018, \mnras, 475, 5659

\bibitem[{{Wright}(2020)}]{wright20}
{Wright} N.~J., 2020, \nar, 90, 101549

\bibitem[{{Wright} {et~al}\mbox{.}(2016){Wright}, {Bouy}, {Drew}, {Sarro},
  {Bertin}, {Cuillandre}, \& {Barrado}}]{wright16}
{Wright} N.~J., {Bouy} H., {Drew} J.~E., {Sarro} L.~M., {Bertin} E.,
  {Cuillandre} J.-C., {Barrado} D., 2016, \mnras, 460, 2593

\bibitem[{{Wright} {et~al}\mbox{.}(2019){Wright}, {Jeffries}, {Jackson},
  {Bayo}, {Bonito}, {Damiani}, {Kalari}, {Lanzafame}, {Pancino}, {Parker},
  {Prisinzano}, {Randich}, {Vink}, {Alfaro}, {Bergemann}, {Franciosini},
  {Gilmore}, {Gonneau}, {Hourihane}, {Jofr{\'e}}, {Koposov}, {Lewis},
  {Magrini}, {Micela}, {Morbidelli}, {Sacco}, {Worley}, \& {Zaggia}}]{wright19}
{Wright} N.~J. {et~al.}, 2019, \mnras, 486, 2477

\bibitem[{{Wright} \& {Mamajek}(2018)}]{wright18}
{Wright} N.~J., {Mamajek} E.~E., 2018, \mnras, 476, 381

\bibitem[{{Wright} {et~al}\mbox{.}(2014){Wright}, {Parker}, {Goodwin}, \&
  {Drake}}]{wright14}
{Wright} N.~J., {Parker} R.~J., {Goodwin} S.~P., {Drake} J.~J., 2014, \mnras,
  438, 639

\bibitem[{{Yuan} {et~al}\mbox{.}(2018){Yuan}, {Chang}, {Banerjee}, {Han},
  {Kang}, \& {Smith}}]{yuan18}
{Yuan} Z., {Chang} J., {Banerjee} P., {Han} J., {Kang} X., {Smith} M.~C., 2018,
  \apj, 863, 26

\bibitem[{{Zari} {et~al}\mbox{.}(2019){Zari}, {Brown}, \& {de Zeeuw}}]{zari19}
{Zari} E., {Brown} A.~G.~A., {de Zeeuw} P.~T., 2019, \aap, 628, A123

\end{thebibliography}
\bsp

\end{document}